\documentclass[twocolumn,aps,superscriptaddress]{revtex4}
\usepackage{graphicx}
\usepackage{float}
\usepackage{dcolumn}
\usepackage{bm}
\usepackage{color}
\usepackage{amsmath}
\usepackage{amssymb}
\usepackage{amsfonts}
\usepackage{esint}
\usepackage{times}
\usepackage{xcolor}
\usepackage{phaistos}
\usepackage{braket}
\usepackage{comment}

\usepackage{pdfpages}

\usepackage[colorlinks,linkcolor=blue,anchorcolor=blue,citecolor=blue]{hyperref}

\begin{document}
\title{Quantum crosstalk analysis for simultaneous gate operations on superconducting qubits}
\author{Peng Zhao}\email{shangniguo@sina.com}
\affiliation{Beijing Academy of Quantum Information Sciences, Beijing 100193, China}
\author{Kehuan Linghu}
\email{linghukh@baqis.ac.cn}
\affiliation{Beijing Academy of Quantum Information Sciences, Beijing 100193, China}
\author{Zhiyuan Li}
\affiliation{Beijing Academy of Quantum Information Sciences, Beijing 100193, China}
\author{Peng Xu}
\affiliation{Institute of Quantum Information and Technology,
Nanjing University of Posts and Telecommunications, Nanjing, Jiangsu 210003, China}
\author{Ruixia Wang}
\affiliation{Beijing Academy of Quantum Information Sciences, Beijing 100193, China}
\author{Guangming Xue}
\email{xuegm@baqis.ac.cn}
\affiliation{Beijing Academy of Quantum Information Sciences, Beijing 100193, China}
\author{Yirong Jin}
\affiliation{Beijing Academy of Quantum Information Sciences, Beijing 100193, China}
\author{Haifeng Yu}
\affiliation{Beijing Academy of Quantum Information Sciences, Beijing 100193, China}

\date{\today}

\begin{abstract}

Maintaining or even improving gate performance with growing numbers of parallel controlled qubits is
a vital requirement for fault-tolerant quantum computing. For superconducting quantum processors,
though isolated one- or two-qubit gates have been demonstrated with high fidelity, implementing these
gates in parallel commonly shows worse performance. Generally, this degradation is
attributed to various crosstalks between qubits, such as quantum crosstalk due to residual inter-qubit coupling. An
understanding of the exact nature of these crosstalks is critical to figuring out respective mitigation
schemes and improved qubit architecture designs with low crosstalk. Here we give a theoretical analysis
of quantum crosstalk impact on simultaneous gate operations in a qubit architecture, where
fixed-frequency transmon qubits are coupled via a tunable bus, and sub-100-ns
controlled-Z (CZ) gates can be realized by applying a baseband flux pulse on the bus. Our analysis
shows that for microwave-driven single-qubit gates, the dressing from the qubit-qubit coupling can cause
non-negligible cross-driving errors when qubits operate
near frequency collision regions. During CZ gate operations, although unwanted nearest-neighbor
interactions are nominally turned off, sub-$\rm MHz$ parasitic next-nearest-neighbor interactions involving
spectator qubits can still exist, causing considerable leakage or control error when one operates qubit
systems around these parasitic resonance points. To ensure high-fidelity simultaneous
operations, there could raise a request to figure out a better
way to balance the gate error from target qubit systems themselves and the error from non-participating
spectator qubits. Overall, our analysis suggests that towards useful
quantum processors, the qubit architecture should be examined carefully in the
context of high-fidelity simultaneous gate operations in a scalable qubit lattice.

\end{abstract}

\maketitle

\section{Introduction}

In pursuit of a useful quantum processor, the quality of qubits (e.g.,
qubit coherence time and fidelity of gate operations), and the number of qubits are
two of the most important figures of merit. Nonetheless, in the practical implementation of
quantum processors, an outstanding challenge is how to maintain or even improve gate
performance with growing numbers of parallel controlled qubits \cite{Martinis2015}. For quantum processors
built with superconducting qubits, isolated single-qubit gates with gate error
below $0.1\%$ \cite{Kjaergaard2020,Chen2016, McKay2017,Somoroff2021} and two-qubit gates with
gate error approaching $0.1\%$ \cite{Kjaergaard2020, Barends2019, Rol2019, Sung2021, Xu2020,Mitchell2021,Wei2021,Sete2021,Ficheux2021}
have been demonstrated in various qubit architectures \cite{Kjaergaard2020}. However, in multi-qubit systems,
implementing gate operations in parallel has commonly shown worse gate performance,
especially for simultaneous two-qubit gate operations applied to nearby qubits,
where gate errors are typically increased by $0.1\%-1\%$ \cite{Chow2015,Arute2019,Zhu2021,Zhang2021,Zajac2021}.
Generally, the performance degradation is caused by the crosstalk effect \cite{Arute2019,Sarovar2020} that
breaks the textbook assumption on quantum computing, i.e., gate operations are implemented spatial
locally and independently (for a high-level and hardware-agnostic definition of
crosstalk, we refer the reader to Ref.$\,$\cite{Sarovar2020} for details). Moreover, this
crosstalk effect can cause not just increased gate errors but even a variety of
correlated and nonlocal errors which are particularly harmful to the realization
of fault-tolerant quantum computing \cite{Fowler2012}. Thus, mitigating the crosstalk effect is critical to
improving gate performance in superconducting quantum processors and achieving
the long-term goal of fault-tolerant quantum computing.

Recently, various studies on the detection, characterization, and mitigation of crosstalk have been
published \cite{Sarovar2020,Gambetta2012,Rudinger2019,Huang2020,Niu2021,Ash-Saki2021,
Rudinger2021,Sung2021,Barends2014,Abrams2019,Dai2021,Takita2016,Takita2016b,McKay2019,
Sundaresan2020,Zajac2021,Winick2021,Deng2021,Tripathi2021,Murali2020,Ding2020,Smith2021}.
From a high-level and hardware-agnostic perspective, several schemes
have been proposed for detecting and characterizing crosstalk in multi-qubit systems \cite{Sarovar2020,Gambetta2012,Rudinger2019,Huang2020,Niu2021,Ash-Saki2021,Rudinger2021}.
Meanwhile, by taking software strategies, various mitigation schemes have also been proposed,
such as, at the qubit- or gate-level, one first characterizes the crosstalk and then actively cancels
it with compensated pulses \cite{Sung2021,Barends2014,Abrams2019,Dai2021,Takita2016,Takita2016b,McKay2019,Sundaresan2020,Zajac2021}
or optimal control pulses \cite{Winick2021,Deng2021,Tripathi2021}, and at the quantum circuit level, one takes better
crosstalk-aware compilation schemes and instruction scheduling schemes for implementing quantum
algorithms \cite{Murali2020,Ding2020,Smith2021}. However, these detection and characterization
schemes cannot in general directly clarify the exact nature of crosstalk, and these software
approaches for mitigating crosstalk commonly require complex control and characterization, thus placing heavy resource
burdens on the control system and limiting their feasibility for large-scale quantum computing.
Given these considerations, besides taking high-level software strategies, it is highly
desirable to further understand crosstalk and figure out how to mitigate crosstalk
from a hardware perspective.

Crosstalk that appears in multi-qubit superconducting quantum processors is commonly very
architecture-dependent. For example, in qubit architectures with fixed coupling and
fixed-frequency qubits \cite{Chow2015}, the dominated crosstalk is associated with microwave
crosstalk and residual inter-qubit coupling \cite{Gambetta2012,Mitchell2021,Wei2021}, whereas in qubit
architectures with frequency-tunable qubits or tunable coupling \cite{Barends2014,Chen2014,Yan2018},
the flux crosstalk is considered as more involved \cite{Barends2014,Abrams2019}. Despite its
architecture-dependent multiformity, from the hardware perspective, crosstalk can be categorized into
two types according to its physical origination \cite{Patterson2019}. One is the classical crosstalk
associated with unintended classical electromagnetic couplings, such as DC or AC flux
crosstalk \cite{Abrams2019,Dai2021} and microwave crosstalk \cite{Sung2021,Mitchell2021,Wei2021}.
The other one is quantum crosstalk resulting from inter-qubit couplings, such as
residual nearest-neighbor (N.N.) coupling \cite{Barends2014,Wei2021} and
next-nearest-neighbor (N.N.N.) coupling \cite{Barends2014,Yan2018,Zhao2020b}. An understanding
of the exact nature of crosstalk or separating error contributions from different
crosstalks is an urgent need for figuring out respective mitigation schemes and for obtaining improved qubit
architecture designs with low crosstalk.  Although several
works have previously examined various quantum crosstalk effects, such as $ZZ$ crosstalk for single-qubit
gates \cite{Gambetta2012,Mundada2019,Li2020}, two-qubit gates \cite{Wei2021,Sundaresan2020,Mundada2019,Ku2020,Zhao2020,Xu2021,Kandala2020,Zhao2021,Finck2021,
Noguchi2020,Mitchell2021,Petrescu2021,Leroux2021,Li2020}, and spectator-qubit induced crosstalk in multi-qubit systems \cite{Takita2016,Takita2016b,McKay2019,Sundaresan2020,Krinner2020,Malekakhlagh2020,Cai2021},
they are restricted mainly to isolated two-qubit gates. Crosstalk analysis in the
context of simultaneous gate operations on a multi-qubit lattice could provide more
physical insight into the nature of the crosstalk effect \cite{Zajac2021,Chu2021} on practically implemented
quantum processors \cite{Arute2019,Zhu2021}.

\begin{figure}[tbp]
\begin{center}
\includegraphics[keepaspectratio=true,width=\columnwidth]{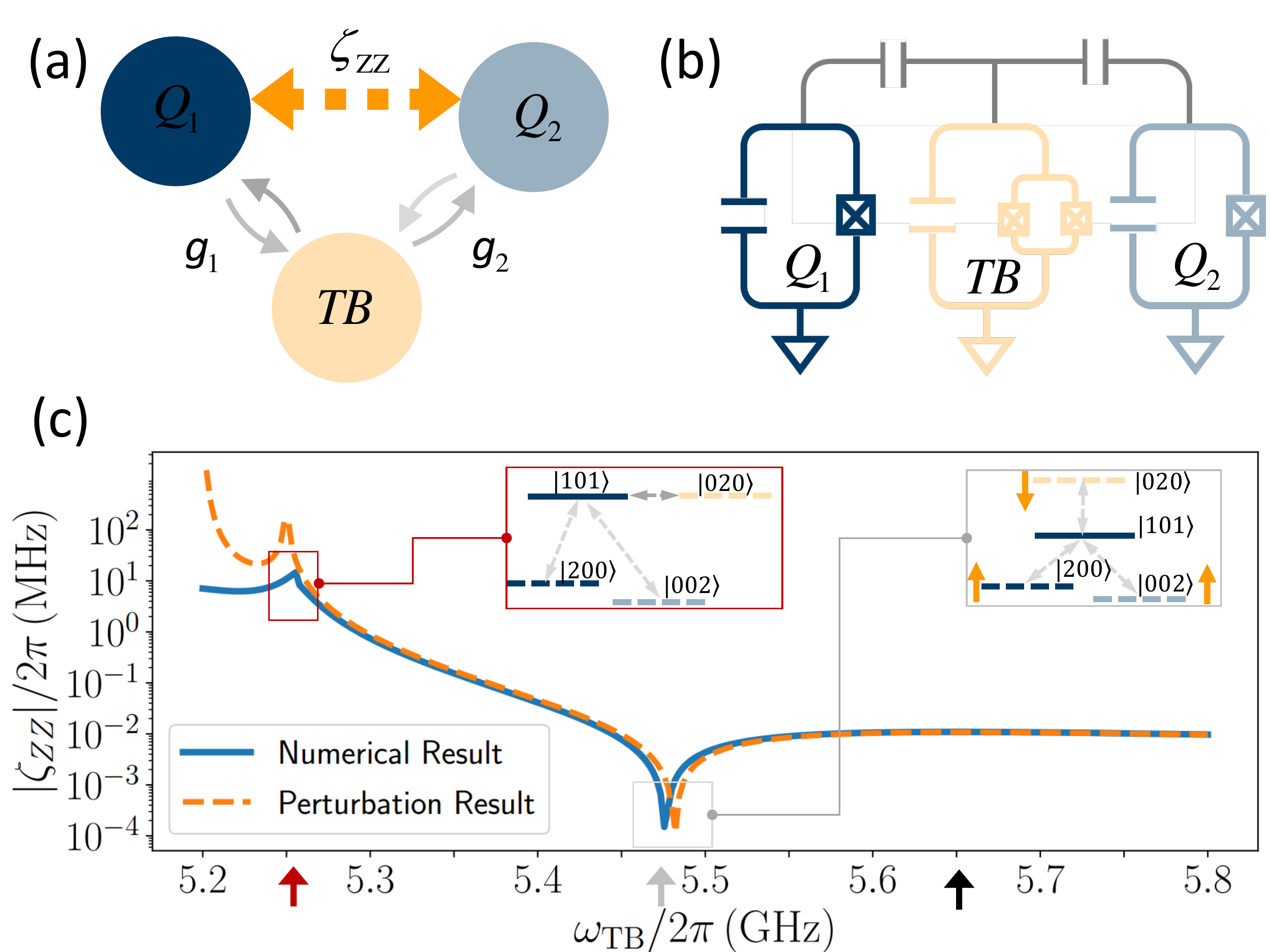}
\end{center}
\caption{(a) Sketch of a coupled two-qubit system ($Q_{1,2}$) with tunable $ZZ$ interaction $\zeta_{\rm ZZ}$
enabled by the tunable bus ($TB$). (b) Circuit diagram of two fixed-frequency transmons coupled via a
frequency-tunable transmon (treated as a tunable bus). (c) $ZZ$ coupling $\zeta_{\rm ZZ}$ as a function
of bus frequency $\omega_{\rm TB}$ for the qubit system illustrated in (b) with qubit frequency
$\omega_{1(2)}/2\pi=5.000(5.200)\,\rm GHz$, anharmonicity $\eta_{1}=\eta_{2}=\eta_{\rm{TB}}=\eta\,(\eta/2\pi=-300\,\rm MHz)$, and
qubit-bus coupling $g_{1(2)}=g\,(g/2\pi=25\,\rm MHz)$ (at $\omega_{1(2)}=\omega_{\rm{TB}}
=5.500\,\rm {GHz})$. The blue-solid and orange-dashed lines represent
the numerical and perturbational results, respectively. The red arrow indicates
the working point where a strong $ZZ$ coupling exists due to the resonance interaction
$|101\rangle\leftrightarrow |020\rangle$ (as shown in the inset outlined
with red line), and grey arrow shows the $ZZ$-free point resulting
from the destructive interference of $ZZ$ contributions from
interactions $|101\rangle\leftrightarrow\{|200\rangle,|002\rangle,|020\rangle\}$ (as
shown in the inset outlined with grey line). To ensure high-fidelity single-qubit addressing, besides the
residual $ZZ$ crosstalk, the cross-driving crosstalk due to qubit state dressing
should also be taken into consideration (see Sec.~\ref{SecIIB} for details). In light
of this, the system idle point is typically $500\,\rm MHz$ above the qubit in frequency,
marked by the black arrow.}
\label{fig1}
\end{figure}

In this work, we give a theoretical analysis of quantum crosstalk impact on simultaneous gate
operations in a qubit architecture with tunable $ZZ$ coupling, as sketched in Fig.~\ref{fig1},
where fixed-frequency transmon qubits \cite{Koch2007} are coupled via a tunable
bus \cite{McKay2016}. To do so, we first give detailed descriptions of the proposed
qubit architecture, especially focusing on the tunable $ZZ$ coupling enabled by the
tunable bus, gate operations, and the scalability of the architecture. We then turn to
illustrate the quantum crosstalk effect on simultaneous gate operations in this qubit
architecture. We consider two typical gate operations, i.e., single-qubit X gate and
two-qubit CZ gate:

(i) For simultaneous single-qubit X gate operations, the dressing from the qubit-qubit
coupling can induce substantial quantum crosstalk, i.e., cross-driving effect, and
the associated crosstalk strength is comparable with that of classical
microwave (see Fig.~\ref{fig4}). Focusing on the nearest-neighbor qubit pairs, we show that
when qubit systems approach the frequency-collision region, the cross-driving effect can make
gate performance worse by almost an order of magnitude (see Fig.~\ref{fig6}). This suggests
that to ensure high-fidelity single-qubit addressing, this dressing-induced cross-driving effect
should be considered as a serious contender of the error sources.

(ii) For simultaneous CZ gate operations, we show that due to the quantum crosstalk, the
performance of simultaneous CZ gate operations can be almost an order of magnitude worse than
that of the isolated case (see Fig.~\ref{fig8}). We further illustrate the exact nature of
these quantum crosstalks, i.e., residual N.N.N. interactions with small coupling
strength, and argue that although unwanted N.N. interactions are nominally turned off,
sub-$\rm MHz$ parasitic N.N.N. interactions involving spectator qubits (enabled by high-order
processes) can still exist. To mitigate their impact on simultaneous two-qubit gate
operations, we show that engineering system parameters (e.g., pushing these
parasitic N.N.N. interaction points away from the system working point) (see Fig.~\ref{fig10}) or implementing
short-time gate operations (see Fig.~\ref{fig12}) could mitigate the quantum crosstalk impact
on simultaneous gate operations. However, on the contrary, gate operations with a slower speed
can, in general, suppress unwanted interactions within target qubit systems. Hence, we argue
that for high-fidelity simultaneous gate operations, this could give rise to a trade-off between
the error resulting from target qubit systems themselves and the error from non-participating
spectator qubits (see Fig.~\ref{fig12}). More strikingly, this trade-off suggests that even
without the consideration of qubit decoherence error, gate operations with slower speeds do not
always show better performance.

Overall, the analysis presented in this work shows that isolated gate performance, especially,
obtained with all non-participating spectator qubits in their ground states, cannot capture
its true performance for the realistic situation in which gates are implemented simultaneously.
Towards a functional quantum processor, the performance of quantum processors should be examined
carefully in the context of simultaneous gate operations. The analysis can improve our understanding
of the exact nature of quantum crosstalk and its impact on simultaneous gate operations, thus may
also pave the way for figuring out crosstalk mitigation schemes and qubit architecture designs with
low crosstalk. Additionally, the result of this analysis also suggests that the proposed qubit
architecture may be a promising architecture towards a large-scale superconducting quantum processor
with low crosstalk.

This paper is organized as follows. In Sec.~\ref{SecII},
we introduce our qubit architecture and give a detailed description of the
gate operation in this qubit architecture. In Sec.~\ref{SecIII},
we numerically study the performance of simultaneous gate operations (i.e., simultaneous
single-qubit and two-qubit gate operations) in the qubit
architecture introduced in Sec.~\ref{SecII}, analyze the leading error
source (i.e., quantum crosstalk) of the added error for the parallel implemented gate
operations, and show how to mitigate quantum crosstalk impact on gate
operations. In Sec.~\ref{SecIV}, we give conclusions of our investigation.

\section{fixed-frequency qubit architectures with tunable buses}\label{SecII}

In this section, we propose a qubit architecture comprising fixed-frequency
transmon qubits coupled via a tunable bus, as shown in Fig.~\ref{fig1}(b). We
first focus specifically on this single building block and show how qubit-bus
coupling can enable a high-contrast $ZZ$ interaction. Therefore, high fidelity single-qubit gates
can be achieved with suppression of $ZZ$ crosstalk, and a sub-100-ns CZ gate can be realized
when we switch on the $ZZ$ interaction between qubits. Then, we consider scaling up
this building block to scalable two-dimensional (2D) qubit grids.

\subsection{System Hamiltonian and tunable ZZ coupling}\label{SecIIA}

Here, we consider a system of two transmon qubits depicted in Fig.~\ref{fig1}(b),
where two transmon qubits $Q_{1,2}$ are coupled through a tunable bus $TB$ (a third ancilla frequency-tunable transmon
qubit). The full system can be modeled by a chain of three
weakly anharmonic oscillators \cite{Koch2007} with nearest-neighboring coupling, and can be
described by (hereafter $\hbar =1$, notation $|Q_{1}\,TB\,Q_{2}\rangle$ represents
states of the full system, and when restricted to the qubit-subspace,
notation $|Q_{1}\,Q_{2}\rangle$ is used),
\begin{eqnarray}
\begin{aligned}\label{eq1}
H=&\sum_{l}\left[\omega_{l}a_{l}^{\dagger}a_{l}+\frac{\eta_{l}}{2}a_{l}^{\dagger}a_{l}^{\dagger}a_{l}a_{l}\right]
\\&+\sum_{n}\left[g_{n}(a_{n}^{\dagger}a_{\rm{TB}}+a_{n}a_{\rm{TB}}^{\dagger})\right],
\end{aligned}
\end{eqnarray}
where the subscript $l\in\{1,2,TB\}$ labels $l$th oscillator $\{Q_{1}\,Q_{2}\,TB\}$ with anharmonicity
$\eta_{l}$ and frequency $\omega_{l}$, $a_{l}\,(a_{l}^{\dagger})$ is
the associated annihilation (creation) operator, and $g_{n}$ ($n=1,2$) denotes the
strength of the coupling between qubit $Q_{n}$ and the tunable bus. The qubit-bus
coupling can mediate an effective $ZZ$ coupling between qubits, which is defined as
\begin{eqnarray}
\begin{aligned}\label{eq2}
\zeta_{\rm ZZ}\equiv(E_{11}-E_{10})-(E_{01}-E_{00})
\end{aligned}
\end{eqnarray}
where $E_{jk}$ denotes eigenenergy of system associated with dressed eigenstate $|\tilde{jk}\rangle$, which is adiabatically
connected to the bare state $|j0k\rangle$. According to the fourth-order perturbation theory \cite{Krishnan1978}, this
effective $ZZ$ coupling can be approximated as \cite{DiCarlo2009,Mundada2019,Zhao2020}
\begin{eqnarray}
\begin{aligned}\label{eq3}
\zeta_{\rm ZZ}=&2g_{1}^{2}g_{2}^{2}\bigg[\frac{1}{\Delta_{2}^{2}(\Delta-\eta_{1})}
-\frac{1}{\Delta_{1}^{2}(\Delta+\eta_{2})}\bigg]
\\&+\frac{2g_{1}^{2}g_{2}^{2}}{\Delta_{1}+\Delta_{2}-\eta_{\rm{TB}}}
\bigg[\frac{1}{\Delta_{1}}+\frac{1}{\Delta_{2}}\bigg]^{2},
\end{aligned}
\end{eqnarray}
where $\Delta_{1(2)}=\omega_{1(2)}-\omega_{\rm{TB}}$, and $\Delta=\omega_{2}-\omega_{1}$ denote the
qubit-bus detuning and qubit-qubit detuning, respectively. In the above expression for $ZZ$
coupling strength, the first two terms result from the interaction
$|101\rangle\leftrightarrow\{|200\rangle,|002\rangle\}$, and the final one contributes
from interaction $|101\rangle\leftrightarrow|020\rangle$ \cite{Zhao2021,Jin2021}.

We first consider how to suppress $ZZ$ coupling at the system idle point, where single-qubit gates
can be implemented with suppression of $ZZ$ crosstalk. From Eq.~(\ref{eq3}), one
can find that $ZZ$-free point, i.e., $\zeta_{\rm ZZ}=0$, is independent of the qubit-bus
coupling strengths. Assuming $\eta_{1}=\eta_{2}=\eta_{\rm{TB}}=\eta$
and considering $\omega_{1}\simeq\omega_{2}$, the $ZZ$-free point is at $\Delta_{1(2)}\simeq3\eta/2$,
where the qubit-bus detuning has a magnitude similar to that of qubit anharmonicity \cite{Goerz2017}.
For transmon qubits with a typical anharmonicity of $300\,\rm MHz$ and operated in the dispersive
regime (i.e., $|\Delta_{1(2)}|\geq 10g_{1(2)}$), this means that to suppress residual $ZZ$ interaction, the qubit-bus
coupling strength should take a value far less than that of the conventional qubit-bus coupling, the
strength of which is typically $100\,\rm MHz$ \cite{McKay2016,Jin2021}. Note here that although qubit systems can in
principle operate in the non-dispersive regime (i.e., $|\Delta_{1(2)}|\ll 10g_{1(2)}$), single
qubit addressing and eliminating unwanted parasitic interactions, such as
N.N.N. coupling, may become highly nontrivial \cite{Goerz2017}.

Thus, below, we consider that the qubit-bus coupling strength takes the value
$g_{1(2)}/2\pi=g/2\pi=25\,\rm MHz$, which is similar to the qubit-qubit coupling strength
in the conventional qubit architecture with fixed direct coupling \cite{Barends2014}.
In this way, at the idle point, where the qubit system operates in the
dispersive regime, both the qubit addressing error and unwanted parasitic interaction
could be suppressed heavily. However, one may doubt that how this rather small
qubit-bus coupling can enable a bus-mediated inter-qubit interaction with adequate
strength for implementing a successful two-qubit gate. Figure~\ref{fig1}(c)
shows the $ZZ$ coupling strength versus bus frequency with
qubit frequency $\omega_{1(2)}/2\pi=5.000(5.200)\,\rm GHz$,
anharmonicity $\eta_{1}=\eta_{2}=\eta_{\rm{TB}}=\eta\,(\eta/2\pi=-300\,\rm MHz)$. The
observation is that the destructive interference of $ZZ$ coupling contributions from
interactions $|101\rangle\leftrightarrow\{|200\rangle,|002\rangle,|020\rangle\}$
(see Fig.~\ref{fig1}(c), the inset outlined with red line) gives rise to a $ZZ$-free point
at frequency $\omega_{\rm{TB}}/2\pi=5.475\,\rm GHz$,
where the qubit-bus detuning is comparable to the qubit anharmonicity. This confirms
the perturbational analysis given above. When biasing the bus at
frequency $\omega_{\rm{TB}}/2\pi\simeq5.250\,\rm GHz$, there exists a $ZZ$ coupling
with a strength of $10\,\rm MHz$, which is adequate and suitable for
implementing a sub-100-ns CZ gate. This $ZZ$ interaction mainly
results from the resonance interaction $|101\rangle\leftrightarrow |020\rangle$
(see Fig.~\ref{fig1}(c), the inset outlined with grey line) at the working
point $\omega_{\rm on}=\bar{\omega}-\eta_{\rm{TB}}/2$,
where $\bar{\omega}=(\omega_{1}+\omega_{2})/2$ denotes
the mean qubit frequency.

\begin{figure}[tbp]
\begin{center}
\includegraphics[keepaspectratio=true,width=\columnwidth]{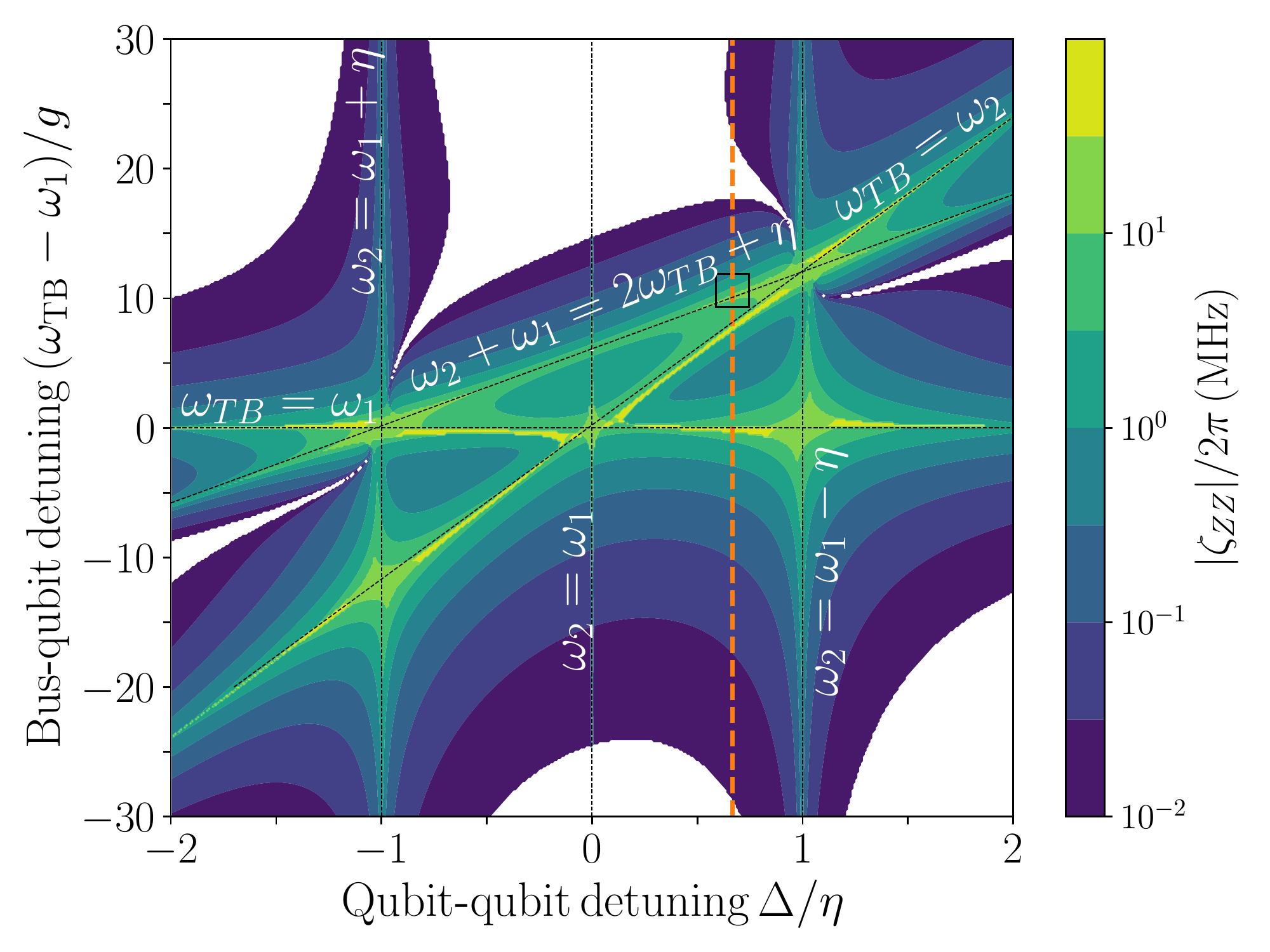}
\end{center}
\caption{$ZZ$ coupling $\zeta_{\rm ZZ}$ (data points with ZZ coupling strengths below 10 kHz are removed)
as a function of qubit-qubit detuning and bus-qubit detuning, where the frequency of $Q_{2}$ is fixed
at $\omega_{1}/2\pi=5.000\,\rm GHz$. Other system parameters are the same as those
used in Fig.~\ref{fig1}(c). Vertical cut (orange dashed line) denotes the result plotted
in Fig.~\ref{fig1}(c), and open black square marks the working point
at $\omega_{\rm on}=\bar{\omega}-\eta_{\rm{TB}}/2$ with the mean qubit
frequency $\bar{\omega}=(\omega_{1}+\omega_{2})/2$.}
\label{fig2}
\end{figure}

In addition, by fixing $Q_{1}$'s frequency at $5.000\,\rm GHz$,
Figure~\ref{fig2} shows the $ZZ$ coupling strength $\zeta_{\rm ZZ}$ (data points with ZZ coupling strengths
below 10 $\rm kHz$ are removed) as a function of qubit-qubit detuning $\Delta$ and bus-qubit
detuning. One can find that suppression of $ZZ$ coupling can
be achieved in several parameter zones, such as the straddling
regime (i.e., qubit-qubit detuning $|\Delta_{1(2)}|<|\eta_{1(2)}|$) \cite{Mundada2019,Goerz2017},
and out of the straddling regime with one qubit above and one qubit below the
bus in frequency. Moreover, contrary to the conventional qubit-bus system \cite{Mundada2019},
one can find that for systems operated out of the straddling regime with qubits below or above the bus
in frequency, $ZZ$ interaction can still be heavily suppressed. This is to be expected, because in the present
qubit system, the qubit-bus coupling is far smaller than that of
the conventional case.

\subsection{Gate operation}\label{SecIIB}

The above illustration shows that the proposed qubit architecture is promising to
achieve high-contrast $ZZ$ interactions in several different parameter regions. Such
flexibility in the choice of parameters can potentially be used to explore various
possible advantages for implementing a scalable quantum processor. In the following
discussion, we focus specifically on the straddling regime, i.e., $|\Delta|<|\eta|$.

\subsubsection{single-qubit gate}\label{SecIIB1}

As usual, the $ZZ$-free point is chosen as the system idle point, thus single-qubit gates
can be implemented without the detrimental effect from residual $ZZ$ interaction. However,
we note that for microwave-activated single-qubit gates, while residual $ZZ$ coupling is
eliminated, there can exist another error source, i.e, cross-driving error. The
cross-driving error describes a phenomenon that microwave driving
applied to one qubit can cause unintended driving on the others \cite{Gambetta2012,Patterson2019}.
This can be due to classical microwave crosstalk or qubit state dressing (resulting from
inter-qubit coupling) induced quantum crosstalk. Here, we restrict ourselves to the last case.
Unlike the cross-driving effect resulting from classical microwave
crosstalk, the qubit-dressing induced one can create a non-local coherent error on
coupled qubits. The exact nature of this cross-driving effect is essentially
the same as the cross-resonance (CR) interaction \cite{Paraoanu2016,Rigetti2010,Gambetta2012},
and CR interaction can be seen as a special cross-driving effect, where one intends
to maximize the "non-local coherent error", thereby, enabling the implementation
of a two-qubit gate, i.e., CR gate.

\begin{figure}[tbp]
\begin{center}
\includegraphics[keepaspectratio=true,width=\columnwidth]{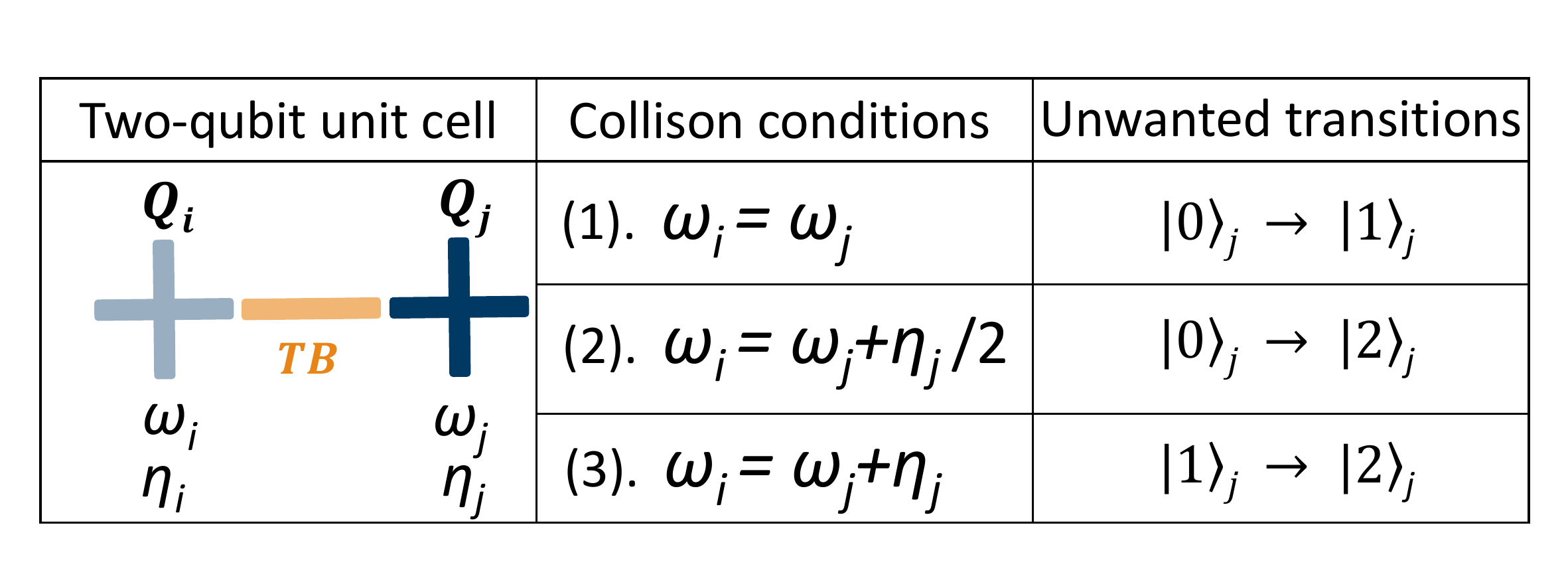}
\end{center}
\caption{Qubit frequency collision conditions and their associated
unwanted transitions.}
\label{fig3}
\end{figure}

Similar to the classical microwave crosstalk, when we focus on an individual qubit,
the dressing-induced cross-driving can excite qubits, causing unintended single-qubit rotation or
leakage out of the qubit-subspace. This can result in a substantially large gate error, especially when
considering the qubit system operated near the frequency collision regions \cite{Malekakhlagh2020,Brink2018}.
As shown in Fig.~\ref{fig3}, there are three main types
of frequency collisions: (Type-1) when the frequencies of two coupled qubits are
close to the on-resonance condition, the cross-driving between qubits can cause transition within
computational subspace, leading to unintended single-qubit rotations; when one qubit's
frequency is nearly on-resonance with the frequency of two-photon transition $|0\rangle\leftrightarrow|2\rangle$ (Type-2) or
high-level transitions $|1\rangle\leftrightarrow|2\rangle$ (Type-3) of others,
the cross-driving can result in population leakage out of qubit-subspace.

\begin{figure*}[tbp]
\begin{center}
\includegraphics[width=16cm,height=9cm]{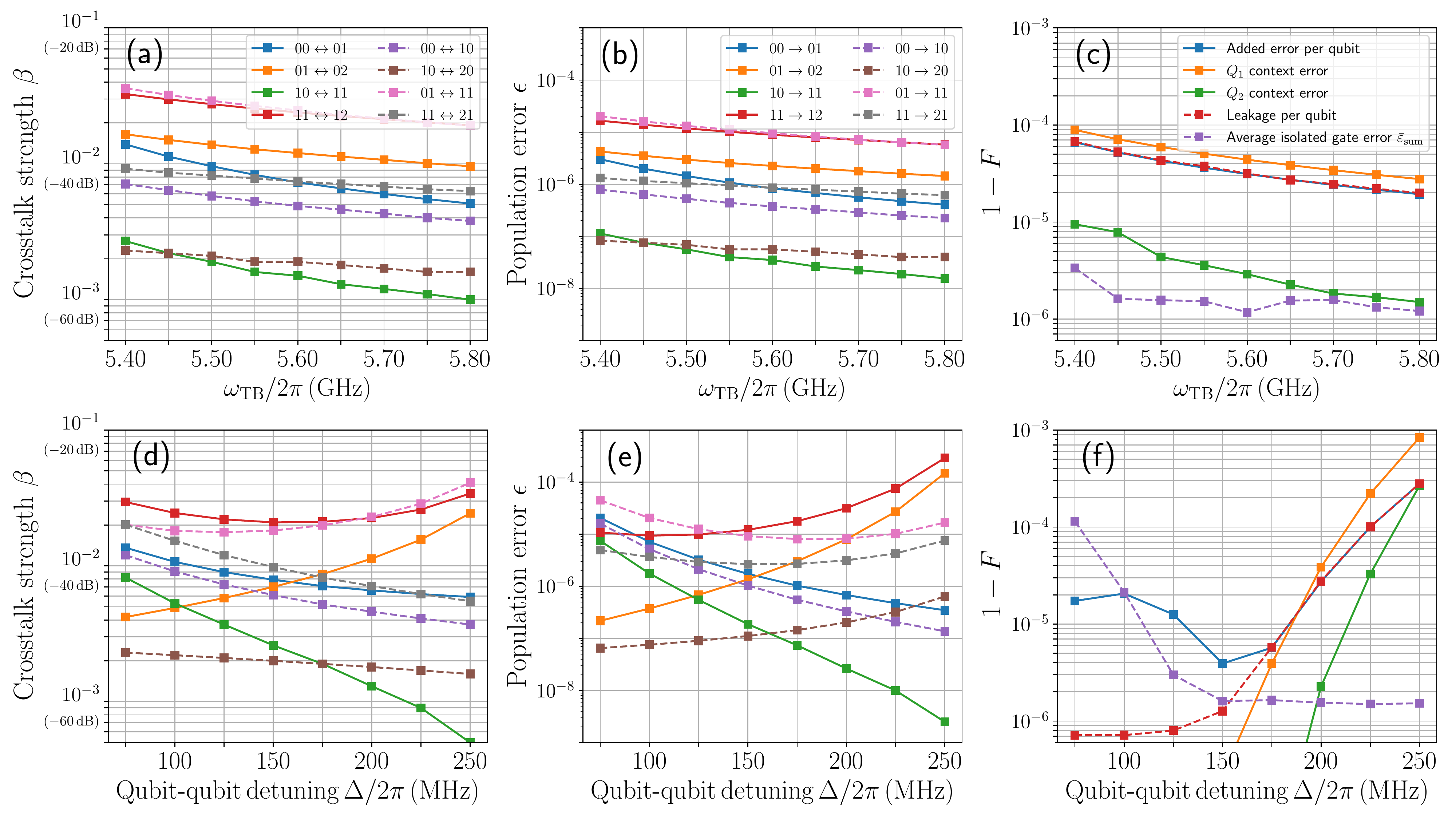}\label{fig5}
\end{center}
\caption{(a-c) Dressing induced cross-driving effect between coupled qubits $Q_{1(2)}$ versus bus
frequency with other system parameters same as those used in Fig.~\ref{fig1}. (d-f) Cross-driving
effect versus qubit-qubit detuning $\Delta$ with bus frequency
$\omega_{\rm TB}/2\pi=5.650\,\rm GHz$ and qubit frequency $\omega_{1}/2\pi=5.000\,\rm GHz$.
Other system parameters are the same as in Fig.~\ref{fig1}. For coupled qubits $Q_{1(2)}$ ,
when a microwave driving with strength $\Omega$ is applied to one qubit, the
nearby qubit can also be felt by driving with
amplitude $\tilde{\Omega}$ (i.e., cross-driving), which is dependent
on qubit states. (a) and (d) show the crosstalk strength between coupled
qubits, defined as $\beta\equiv\tilde{\Omega}/\Omega$. Here, $0j\leftrightarrow 0k$
and $1j\leftrightarrow 1k$ label the cross-driving applied on the transition
$|j\rangle\leftrightarrow|k\rangle$ of $Q_{2}$ with $Q_{1}$ in its ground state and excited
state, respectively. Similarly, $j0\leftrightarrow k0$ and $j1\leftrightarrow k1$
label the cross-driving applied on the transition
$|j\rangle\leftrightarrow|k\rangle$ of $Q_{1}$ with $Q_{2}$ in its ground state and excited
state, respectively. (b) and (e) show the worst-case
population error caused by cross-driving. Here, $0j\rightarrow 0k$ ( $1j\rightarrow 1k$) labels the
worst case population on $|k\rangle$ of $Q_{2}$, where $Q_{2}$ is prepared in
state $|j\rangle$ and the $Q_{1}$ is in its ground state (excited state). (c) and (f) present
the single-qubit gate error ($1-F$) under the influence of the cross-driving effect between qubits.}
\label{fig4}
\end{figure*}

For illustration purposes, here we consider a microwave driving with a constant
amplitude applied to $Q_{1}$ in the two-qubit system described
by Hamiltonian in Eq.~(\ref{eq1}), intending to implement a single-qubit gate on $Q_{1}$.
The driven Hamiltonian can be described by
\begin{eqnarray}
\begin{aligned}\label{eq4}
H_{d}=[\Omega_{x}(t)\cos(\omega_{d}t)+\Omega_{y}(t)\sin(\omega_{d}t)](a_{1}^{\dagger}+a_{1}),
\end{aligned}
\end{eqnarray}
with $\Omega_{d}(t)\equiv\Omega_{x}(t)+i\Omega_{y}(t)$, where $\Omega_{d}(t)=\Omega$
denotes the driving amplitude, and $\omega_{d}$ is the driving frequency.
Similar to the CR interaction, the cross-driving
from $Q_{1}$ to $Q_{2}$ is dependent on the state of $Q_{1}$. By diagonalizing
the undriven system Hamiltonian (at the idle point) in Eq.~(\ref{eq1}),
one can obtain the dressed eigenstate $|\tilde{jk}\rangle$. Then, rewriting the driven
Hamiltonian Eq.~(\ref{eq3}) in terms of the dressed eigenstate, one can obtain strength
of the cross-driving from $Q_{1}$ to $Q_{2}$ (similar analysis can also be applied to the cross-driving
from $Q_{2}$ to $Q_{1}$). For example, the strength of the cross-driving from $Q_{1}$ to $Q_{2}$
can be approximated as
\begin{eqnarray}
\begin{aligned}\label{eq5}
&\tilde{\Omega}_{00\leftrightarrow01}\simeq\frac{J\Omega}{\Delta},
\\&\tilde{\Omega}_{10\leftrightarrow11}\simeq\frac{J\Omega(\Delta+\eta_{1})}{\Delta(\Delta-\eta_{1})},
\\&\tilde{\Omega}_{01\leftrightarrow02}\simeq \frac{\sqrt{2}J\Omega}{\Delta+\eta_{2}},
\\&\tilde{\Omega}_{11\leftrightarrow12}\simeq\frac{\sqrt{2}J\Omega(\Delta+\eta_{1}+\eta_{2})}{(\Delta-\eta_{1}+\eta_{2})(\Delta+\eta_{2})},
\end{aligned}
\end{eqnarray}
where $\tilde{\Omega}_{0j\leftrightarrow0k}$ and $\tilde{\Omega}_{1j\leftrightarrow1k}$
denote the cross-driving applied to $Q_{2}$'s transitions $|j\rangle\leftrightarrow|k\rangle$
for $Q_{1}$ in the ground and excited states, respectively\cite{Magesan2020},
and $J$ denotes the effective bus-mediated exchange coupling, which can be approximated
as $J=(g_{1}g_{2}/2)(1/\Delta_{1}+1/\Delta_{2})$.

Similar to the characterization of classical microwave crosstalk, we can define the crosstalk
strength of this dressing-induced cross-driving as $\beta\equiv\tilde{\Omega}/\Omega$.
Moreover, to quantify the cross-driving induced gate error, here we also introduce the
worst-case population error (since off-resonance driving can cause incomplete Rabi
oscillation, here we choose the maximum population amplitude as an indicator of the gate error and
set $\Omega/2\pi=25\,\rm MHz$) defined as
\begin{eqnarray}
\begin{aligned}\label{eq6}
\epsilon\equiv\frac{\tilde{\Omega}^{2}}{\tilde{\Omega}^{2}+\delta^{2}},
\end{aligned}
\end{eqnarray}
where $\delta$ denotes the detuning between the cross-driving frequency and
qubit's transition frequencies between different energy levels.

To ensure high-fidelity single-qubit gates, the cross-driving effect should be minimized.
From the expression given in Eq.~(\ref{eq5}), this can be achieved by reducing
exchange coupling $J$ and avoiding the qubit frequency collisions
shown in Fig.~\ref{fig3}. Figures~\ref{fig4}(a) and~\ref{fig4}(b) show the crosstalk
strength $\beta$ between $Q_{1}$ and $Q_{2}$, and the cross-driving induced population error
versus bus frequency. One can find that both the crosstalk strength and the population error
are indeed decreased as we increase the bus frequency.
In Figs.~\ref{fig4}(d) and~\ref{fig4}(e), we also show the crosstalk strength and the
population error versus qubit-qubit detuning $\Delta$. We find that the crosstalk strength
itself is dependent on the detuning $\Delta$, and it is also highly asymmetric
(also see Fig.~\ref{fig4}(a)). For example, for qubits with lower
frequency, i.e., $Q_{1}$, the cross-driving from $Q_{2}$ (with a higher qubit frequency)
to $Q_{1}$ becomes notable when $Q_{2}$ is in the excited state
(e.g., $|\tilde{01}\rangle\leftrightarrow|\tilde{11}\rangle$). Meanwhile, the
crosstalk from $Q_{1}$ to $Q_{2}$ especially tends to give leakage out
of qubit subspace (e.g., $|\tilde{11}\rangle\leftrightarrow|\tilde{12}\rangle$
and $|\tilde{01}\rangle\leftrightarrow|\tilde{02}\rangle$). Nevertheless,
as shown in Fig.~\ref{fig4}(e), on the whole, the cross-driving impact on
gate fidelity can still be captured by the frequency collision analysis:

(i) For coupled qubits with small qubit-qubit detuning, the system approaches
Type-1 frequency collision, thus the dominated error from cross-driving
is unintended single-qubit rotations, whereas the leakage error is
suppressed.

(ii) When qubit-qubit detuning approaches the anharmonicity of qubits,
i.e., the Type-3 frequency collision, the cross-driving induced leakage error is
dominated and unintended rotation is greatly reduced. This
suggests a fundamental trade-off between control error and leakage
error.

(iii) The Type-2 frequency collision seems to have less
impact on gate performance \cite{Kelly2014}. This is reasonable since the Type-2
frequency collision involves a second-order process that tends to be orders
of magnitude weaker than that of Type-1 and Type-3.

In the discussion given above, the population error is just a rough
estimation of the cross-driving induced gate error \cite{Malekakhlagh2021}.
In practical implementations of single-qubit gates, rather than using a constant
magnitude driving, one commonly uses a shaped pulse, e.g.,
Derivative Removal by Adiabatic Gate (DRAG) pulse \cite{Motzoi2009,Chen2016,McKay2017},
thus suppressing leakage out of qubit-subspace.
In the following discussion, to implement a single qubit $X$ gate, we consider using DRAG
scheme with a 20-ns cosine-shaped pulse \cite{Chen2016}, given as $\Omega_{\rm DRAG}(t)=
\Omega_{x}(t)-i\alpha\dot{\Omega}_{x}(t)/\eta$, where
$\Omega_{x}(t)=A[1-\cos(2\pi t/t_{g})]/2$, $t_{g}$ is the gate time,
$A$ is the peak pulse amplitude (for $t_{g}=20\,\rm ns$, $A/2\pi\simeq50\,\rm MHz$), $\alpha$
is a free parameter that is chosen
to suppress leakage error (see Appendix~\ref{A1} for details
on the calibration procedure of X gates) \cite{Motzoi2009,Chen2016,McKay2017}.

To give a more exact characterization of the cross-driving
impact on single-qubit gates, we consider the following virtual
experiment: (1) First, an isolated single-qubit gate on qubit $Q_{1(2)}$ is tuned up and
characterized with the other qubit in its ground state, giving rise to the isolated
gate fidelity, i.e., $F_{\blacksquare0}$ (here the subscript denotes the state bitstring
$|\blacksquare 0\rangle$, and for the active target system to which a gate operation
is applied, its qubit state is labeled by $\blacksquare$) for $Q_{1}$ and $F_{0\blacksquare}$ for $Q_{2}$.
(2) Then, the same gate is re-characterized with the other qubit in
its excited state, giving rise to $F_{\blacksquare1}$ for $Q_{1}$ and $F_{1\blacksquare}$
for $Q_{2}$. (3) Finally, we consider that both gates are implemented simultaneously, characterization of
this composite gate, i.e., $\rm{X\otimes X}$, gives rise to the gate
fidelity $F_{\rm{X\otimes X}}$. Here, we define $Q_{1}$ ($Q_{2}$) context
error $|F_{\blacksquare 0}-F_{\blacksquare 1}|$ ($|F_{0\blacksquare}-F_{1\blacksquare}|$)
to characterize the cross-driving impact on isolated single-qubit gates, and define added error per qubit $(F_{\blacksquare0}F_{0\blacksquare}-F_{\rm{X\otimes X}})/2$ for simultaneous single-qubit
gate operations.

To quantify the effect of cross-driving on the gate performance, we use
the metric of gate fidelity given as (without qubit decoherence) \cite{Pedersen2007}
\begin{eqnarray}
\begin{aligned}\label{eq7}
F=\frac{{\rm Tr}(\tilde{U}^{\dagger}\tilde{U})+|{\rm Tr}(U^{\dagger}\tilde{U})|^{2}}{d(d+1)},
\end{aligned}
\end{eqnarray}
where $U$ and $\tilde{U}$ denote the target gate operation and implemented gate
operation, respectively, and $d$ represents the system dimension. In Figs.~\ref{fig4}(c) and~\ref{fig4}(f),
we show the fidelity of $X$ gate as a function of bus frequency and qubi-qubit detuning, respectively.
Similar to the result shown in Figs.~\ref{fig4}(b) and~\ref{fig4}(e), one can find
that reducing exchange coupling $J$ indeed suppresses the cross-driving effect on single-qubit gates,
and a non-negligible gate error of about $0.002\%-0.1\%$ (given the state-of-the-art
single-qubit gate error approaching $0.01\%$ \cite{Chen2016, McKay2017}) appears
when the qubit-qubit detuning approaching Type-1 or Type-3 frequency collision region.
To identify the nature of the added error, we also show the added leakage error per
qubit \cite{Wood2018}. As shown in Fig.~\ref{fig4}(f), for qubit-qubit detuning near
the Type-3 frequency collision, the leakage error is indeed the dominated error
source, accounting for the added error during simultaneous gate operations.
Moreover, the context error also coincides with the leakage, this is true since
leakage out of qubit-subspace can only occur when the cross-driven qubit is in
the excited state. On the contrary, the context error and leakage errors are far
less important near the Type-1 frequency collision region, in which the dominated
error is the unintended single-qubit rotations, e.g., under- or over-rotation (bit-flip error) and
an additional single-qubit phase error (phase error) when one focuses on
each individual qubit.

In addition, it is worth mentioning that under our definition of gate
characterization, we always assume that the other nearby qubit is in the ground
or excited state, thereby, the cross-driving induced flip of nearby qubit
states contributes to errors of the isolated single-qubit gate, i.e., leading to
leakage error. Therefore, the added error per qubit cannot capture the cross-driving induced
single-qubit rotation error. As shown in Fig.~\ref{fig4}(f), near the
Type-1 frequency collision region, although the added error per qubit is rather small,
the averaged isolated single-qubit error
$\bar{\varepsilon}_{\rm {sum}}=(F_{\blacksquare0}+F_{0\blacksquare})/2$ is far larger
than that for Type-3 frequency region. As we mentioned before, when the qubit system works nearby
the Type-1 frequency collision region, the cross-driving effect can lead to unintended
single-qubit rotations on one qubit when single-qubit operations are applied to its
neighboring qubits. This explains the increase in average gate error $\bar{\varepsilon}_{\rm {sum}}$
when decreasing qubit-qubit detuning. We thus argue that restricted totally to the isolated
single-qubit system regardless of nearby qubits states, i.e., tracing out nearby qubits,
one can only achieve a nominally high-fidelity single-qubit gate, which cannot account
for its practical performance on multi-qubit systems.

Overall, for the present system operated in the straddling regime, the dressing
from the qubit-qubit coupling can induce a substantial cross-driving
effect, and the associated crosstalk strength (as shown in Figs.~\ref{fig4}(c) and~\ref{fig4}(d), typically,
from $-40\,\rm dB$ to $-20\,\rm dB$) is comparable with that of classical microwave
crosstalk \cite{Patterson2019,Huang2021}. For systems with
small qubit-qubit detuning (i.e., near the Type-1 frequency collision region), cross-driving favors bit-flip error or phase error,
whereas, for large detuning cases (i.e., near the Type-3 frequency collision region), cross-driving tends to cause leakage error.
To ensure high-fidelity single-qubit addressing, the bus-qubit detuning should
be large enough, thus bus-mediated exchange coupling is suppressed, and the
qubit-qubit detuning should also be away from the frequency collision regions,
especially for the Type-1 and Type-3 \cite{Kelly2014}. Note here that in principle,
the direct coupling between qubit and bus can also cause cross-driving from
qubit to bus. However, according to Eqs.~(\ref{eq5}) and~(\ref{eq6}), one can
find $\epsilon\propto1/\Delta^{3}$, thus, here, the cross-driving effect on the
bus is heavily suppressed by the large qubit-bus detuning. This was checked for the
present system and was found to be less important.

\begin{figure*}[tbp]
\begin{center}
\includegraphics[width=16cm,height=6cm]{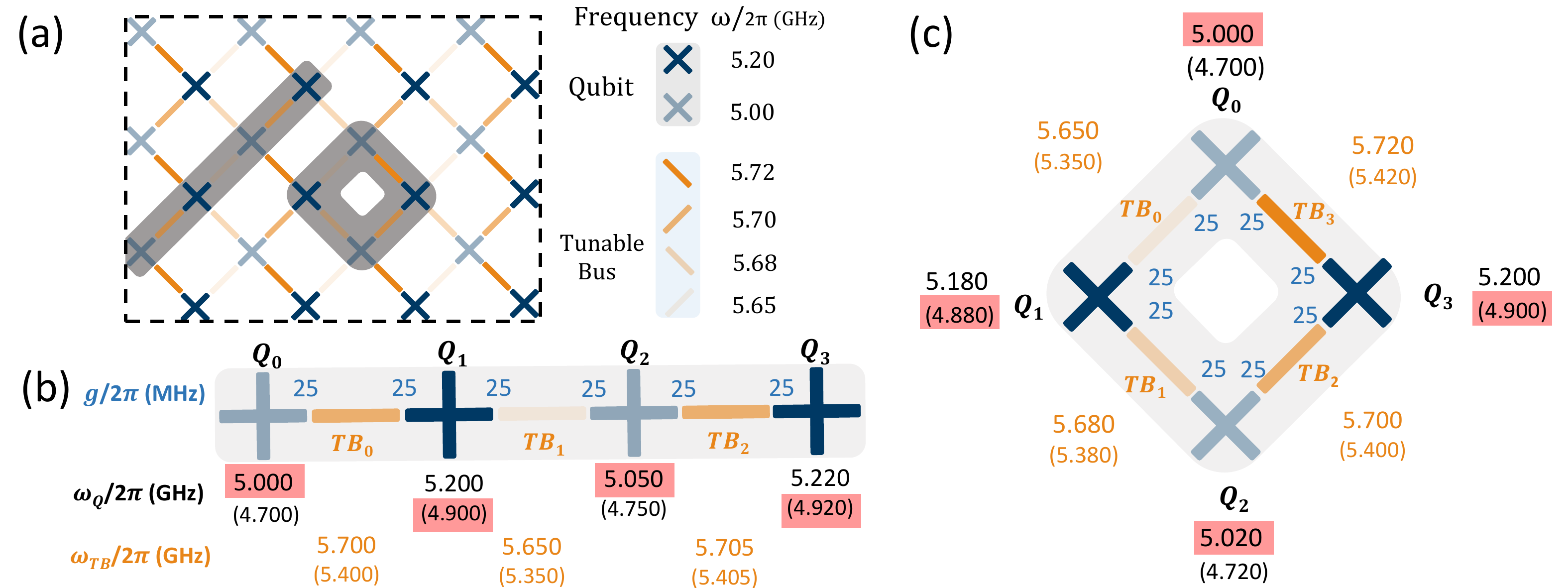}
\end{center}
\caption{(a) Layout of a two-dimensional qubit lattice, where cross-shaped vertices (steel-blue)
denote qubits, and strip-shaped edges (orange) between adjacent qubits represent tunable
buses. Qubits in the lattice belong to one of the two frequency bands, i.e., a lower frequency band with
typical frequency $\omega/2\pi=5.000\,\rm GHz$ (light steel-blue)  or a higher frequency band with
typical frequency $\omega/2\pi=5.200\,\rm GHz$ (dark steel-blue). The typical detuning between nearby
qubits is about $200\,\rm MHz$, and at the idle point, the bus frequency is typically above $500\,\rm MHz$
of the qubits in the higher frequency band. Example of system parameters of four-qubit chain
and $2\times2$ four-qubit square shown in the grey shadow of (a) is presented in (b) and (c),
respectively, e.g., for $Q_{0}$ in the four-qubit chain shown in (b), the qubit frequency is
$\omega_{Q}/2\pi=5.000\,\rm GHz$, and its anharmonicity is $-300\,\rm MHz$ (i.e., its
transitions frequency between $|1\rangle$ and $|2\rangle$ is $4.700\,\rm GHz$), the
interaction between qubit $Q_{0}$ and bus $TB_{0}$ is $g/2\pi=25\,\rm MHz$ (at
$\omega_{Q}/2\pi=\omega_{TB}/2\pi=5.500\,\rm GHz$ ). }
\label{fig5}
\end{figure*}

\subsubsection{two-qubit gate}\label{SecIIB2}

As demonstrated in Sec.~\ref{SecIIA}, for the present qubit architecture
operated in the straddling regime, a high-contrast control over $ZZ$ interaction
can be achieved. When tuning the system energy level $|101\rangle$
on-resonance with $|020\rangle$, i.e., biasing the bus to the working point
at $\omega_{\rm on}=\bar{\omega}-\eta_{\rm{TB}}/2$, a sub-100-ns CZ gate
can be achieved by using diabatic scheme \cite{Barends2019,Linghu2020}
or fast-adiabatic scheme \cite{Xu2020,Martinis2014,Collodo2020,Stehlik2021}.
In principle, during the idle time, the system (tunable bus) can be biased to the exact $ZZ$-free
point. However, as discussed in Sec.~\ref{SecIIB1},
to have a high fidelity single-qubit addressing, besides residual $ZZ$
crosstalk, the cross-driving effect should also be minimized. Thus, here the system idle
point is chosen so that the typical bus frequency
is $500\,\rm MHz$ above the qubit in frequency. In this way, the residual $ZZ$ coupling
can still be suppressed below $10\,\rm kHz$, as shown in Figs.~\ref{fig1}(c) (marked
by the black arrow) and~\ref{fig2}.
Moreover, to avoid the harmful frequency collision in the straddling regime,
the qubit-qubit detuning should be located within
specific limits, e.g., $100\,\rm MHz\,-\,200\,\rm MHz$, thereby, the
single-qubit gate error can be pushed below $0.01\%$ (or even
approaches $0.001\%$), as shown in Fig.~\ref{fig4}(f).

In the present qubit architecture, to implement a successful two-qubit gate,
several practical experimental limitations also need to be considered seriously:

(1) Uncertainty of system parameters due to fabrication. This is one of the most critical
issues faced by fixed-frequency qubits. As shown in Fig.~\ref{fig2},
both the available $ZZ$-free zone (where the $ZZ$ coupling strength is
below 10 $\rm kHz$), and the available working zone (where resonance
interaction $|101\rangle\leftrightarrow |020\rangle$
gives rise to a $ZZ$ coupling with strength about $10\,\rm MHz$, indicated
by black-dashed line labeled by $\omega_{1}+\omega_{2}=2\omega_{\rm TB}+\eta$)
in the straddling regime show a promising insensitivity against variation
of qubit-qubit detuning and qubit-bus detuning. Thus, the parameter uncertainty
issues could be largely fixed by combing this parameter insensitivity and
the state-of-the-art fabrication techniques \cite{Kreikebaum2020,Hertzberg2020}.

(2) Gate error resulting from qubit relaxation and dephasing. Although
fixed-frequency transmon qubit itself can have a long coherence time, qubit relaxation
rate or dephasing rate can be enhanced through the tunable bus, which is more vulnerable to
system noise, especially, $1/f$ flux noise. Moreover, during the CZ gate operation,
a baseband flux pulse is commonly applied to the tunable bus, tuning its frequency
from the idle point to the working point. Considering the qubit system shown in
Fig.~\ref{fig1}, at the working point, the qubit-bus detunings are
$\Delta_{1}/2\pi=-250\,\rm MHz$ and $\Delta_{2}/2\pi=-50\,\rm MHz$.
We expect that the small qubit-bus detuning at the working point can cause a strong hybrid
between qubit and bus (qubit-bus dressing), especially for $Q_{2}$, thus the qubit relaxation
and dephasing through the bus can become greatly serious \cite{McKay2016,Xu2020,Collodo2020,Stehlik2021}.
For the $1/f$ noise induced dephasing, this can be mitigated by employing a tunable bus, whose maximum frequency
point (i.e., flux-insensitive point) is around the idle point. Therefore, at
the idle point, the qubit dephasing through the bus is reduced. Moreover, during CZ
gate operations, the needed frequency excursion of the bus is smaller than $500\,\rm MHz$,
thereby potentially greatly reducing flux sensitivity of bus and mitigating the effect of $1/f$
flux noise \cite{Arute2019,Foxen2020} (in principle, this could be further improved by using tunable buses
with asymmetric SQUID \cite{Hutchings2017}).

(3) Flux pulse distortion. Flux pulse distortion is a critical problem
faced by baseband flux controlled gate operations. The presence of flux pulse
distortion can cause a substantial gate error, such as leakage error and gate bleedthrough,
during CZ gate operations \cite{Barends2014,Foxen2020,Martinis2014,Kelly2014}. Although
pulse distortion can be largely corrected, one needs to characterize
it at first \cite{Barends2014}. Unfortunately, in general, one can not directly characterize this
distortion without control and measurement of the system subjected to pulse distortion. In
qubit architectures with tunable buses or couplers, the bus or coupler, in general, does not
have a dedicated driving line and readout resonator. Still, one can put these back,
but it will greatly increase circuit complexity \cite{Sung2021}. Commonly,
due to the presence of (classical or dressed-induced) microwave crosstalk, the bus or
coupler can be addressed through the driveline of nearby qubits. However, the readout of
bus or coupler state through the readout resonator of nearby qubits is still a challenge.
One possible strategy is that combined with the readout of nearby qubits, full control
on the bus could be achieved through the qubit-bus interaction
\cite{Leek2009,Caldwell2018,Pechal2020}, then the distortion could be
corrected \cite{Jerger2019,Rol2020,Barends2014}. Meanwhile, in the present qubit
architecture, the frequency excursion of the bus is small, thereby potentially
mitigating the effect of flux setting tails caused by pulse distortion \cite{Foxen2020,Foxen2019}.

In the present work, we consider that during the CZ gate operation, a fast adiabatic pulse is
applied to the tunable bus, tuning its frequency from the idle point ($\theta_{i}$) to the
working point $\omega_{\rm{on}}$ ($\theta_{f}$) and then back.
Expressed in terms of Fourier basis functions, the pulse shape is described as \cite{Martinis2014}

\begin{eqnarray}
\begin{aligned}\label{eq8}
\theta(t)=\theta_{i}+ \frac{\theta_{f}-\theta_{i}}{2}\sum\limits_{n=1,2,3...}\lambda_{n}\left[1-\cos\frac{2n\pi t}{T}\right]
\end{aligned}
\end{eqnarray}
with constraints on the odd coefficients $\Sigma_{n\,\rm{odd}}\,\,\lambda_{n}=1$,
and control angle $\theta\equiv\arctan(2J_{101}/\Delta_{101})$, where $J_{101}$ denotes the strength of resonance
interaction $|101\rangle\leftrightarrow|020\rangle$, $\Delta_{101}$
represents the detuning between $|101\rangle$ and $|020\rangle$, and $T$ is the gate time.

\subsection{Scaling up to two-dimensional qubit grids}\label{SecIIC}

In principle, the above two transmon qubits system can be exploited as
building blocks of scalable qubit architectures. An illustration of a
2D qubit lattice, where each transmon qubit is coupled to four nearest
neighbors through tunable buses, is shown in Fig.~\ref{fig5}(a). In the 2D qubit lattice,
the allocation of qubit and bus frequency follows the following rules:

(1) The frequency of the qubit in each row and
column belongs to one of the two frequency bands, i.e., a lower frequency band with
typical frequency $\omega/2\pi=5.000\,\rm GHz$ and a higher frequency band with
typical frequency $\omega/2\pi=5.200\,\rm GHz$, and qubits in the nearby row or column belong to
different frequency bands. Thus, typical detuning between N.N. qubits is
about $200\,\rm MHz$, which is adequate to mitigate cross-driving effect at idle
point (Sec.~\ref{SecIIB}) and to implement a sub-100-ns CZ gate, as demonstrated
in Sec.~\ref{SecIIA}. Moreover, to mitigate microwave crosstalk impact
on qubit addressing between next-nearest-neighbor qubits \cite{Barends2014,Kelly2014}
at the idle point and suppress N.N.N. coupling
between qubits at the idle and the working point, frequency degeneracy in the
lower- or higher-frequency band breaks, yielding typical qubit detuning between
next-nearest-neighbors of about $20\,\rm MHz$.

(2) To have a better balance between the requirement for suppressing qubit-qubit residual
interaction and to limit the frequency excursion range during CZ gate operations, the bus
frequency is typically above $500\,\rm MHz$ of the qubit in the higher frequency band, as
shown in Fig.~\ref{fig5}(a).

When considering a practical realization of this qubit lattice, frequency uncertainty due to
fabrication is the most critical issue, as discussed in Sec.\ref{SecIIB2}.
Besides striving to improve qubit device fabrication precision \cite{Kreikebaum2020,Hertzberg2020},
one could also try to adopt more sparse qubit connections towards large-scale qubit
lattice \cite{Hertzberg2020}.

\section{quantum crosstalk analysis for simultaneous gate operations}\label{SecIII}

In this section, we give a quantum crosstalk analysis for simultaneous
gate operations on two four-qubit systems shown in Figs.~\ref{fig5}(b)
and~\ref{fig5}(c), where system-specific parameters are also presented. Similar to the
discussion in Sec.~\ref{SecIIB1}, to characterize the quantum crosstalk effect
on gate performance, we use the metric of gate fidelity given in Eq.~(\ref{eq7}).
The isolated gate fidelity is obtained by characterizing each gate operation
individually with all other non-participating spectator qubits at their ground states,
and the simultaneous gate fidelity is obtained by characterizing the parallel
implemented gate operations (see Appendix~\ref{A} for details on the calibration procedure). To have enough sensitivity to
the added error caused by quantum crosstalk during simultaneous gate operations, each
isolated gate is optimized numerically, giving typically isolated
single-qubit gate error at $10^{-6}$ and typical two-qubit CZ gate error
at $10^{-5}$. Note here that we don't attempt to search for an optimal
isolated gate operation with the lowest error, but only for gate operations with
enough fidelity, thus enabling an adequate crosstalk sensitivity.

In addition, we note here that in the following discussion,
notations $|Q_{0}\,Q_{1}\,Q_{2}\,Q_{3}, TB_{0}\,TB_{1}\,TB_{2}\,TB_{3}\rangle$
and $|Q_{0}\,Q_{1}\,Q_{2}\,Q_{3}, TB_{0}\,TB_{1}\,TB_{2}\rangle$ represent
states of the four-qubit square and four-qubit chain system, respectively. When
confined to qubit subspace, notation $|Q_{0}\,Q_{1}\,Q_{2}\,Q_{3}\rangle$ is used.

\subsection{Simultaneous single-qubit X gate operations}\label{SecIIIA}
\begin{figure}[tbp]
\begin{center}
\includegraphics[keepaspectratio=true,width=\columnwidth]{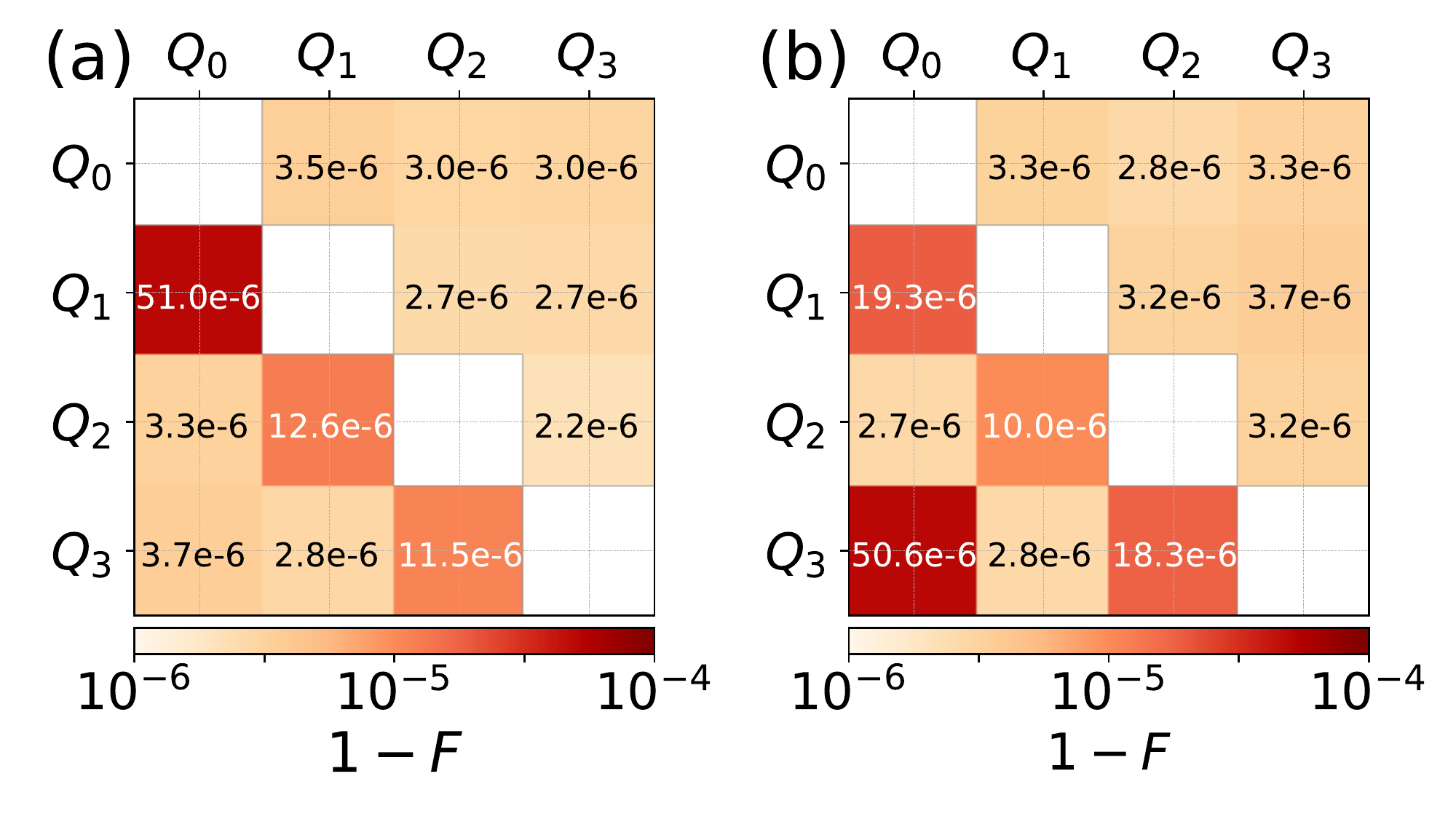}
\end{center}
\caption{Gate error of simultaneous single-qubit $X$ gates (i.e., composite gate $\rm X\otimes X$) applied to arbitrary qubit-pair of (a) the
four-qubit chain system and (b) the $2\times2$ four-qubit square system shown in
Fig.~\ref{fig5}(b) and ~\ref{fig5}(c). Elements in the upper triangular part of the error
matrix denote ideal composite gate error obtained by adding the errors of constituted isolated
gates, and elements in the lower triangular parts correspond to the gate error of
simultaneous gate operations.}
\label{fig6}
\end{figure}

Here, we consider that two 20-ns single-qubit $X$ gates are applied simultaneously to
the arbitrary qubit-pair of the four-qubit system, i.e., implementing a composite X
gate ($X\otimes X$). All tunable buses are tuned to their idle points, thus
residual $ZZ$ interaction between qubits is suppressed below $10\,\rm kHz$. Same as the
discussion in Sec.~\ref{SecIIB1}, single-qubit gates are realized by using the DRAG scheme with
a cosine-shaped pulse.

Figures~\ref{fig6}(a) and~\ref{fig6}(b) show the error matrix of the parallel implemented X gates in the four-qubit
chain system shown in Fig.~\ref{fig5}(b) and the four-qubit square system shown in
Fig.~\ref{fig5}(c), respectively. Elements in the upper and lower triangular part of the error
matrix denote ideal composite gate error obtained by adding the errors of constituted isolated
gates and directly characterizing the parallel implemented gate operations, respectively. We find
that almost no additional error is observed for the next-nearest-neighbor qubits, and we thus conclude that
the quantum crosstalk between next-nearest-neighbor qubits can be neglected. However, focusing on the
nearest-neighbor qubit-pairs, the performance of simultaneous gate operations is almost an order of
magnitude worse than that of the isolated case.

The above observations are consistent with the analysis of dressing induced cross-driving
effect on the gate performance discussed in Sec.~\ref{SecIIB1}. As interactions
between next-nearest-neighbor qubits are commonly an order of magnitude weaker than
that of nearest-neighbor qubits, the cross-driving effect can be greatly suppressed.
On the contrary, the cross-driving effect between nearest-neighbor qubits cannot be ignored,
and in general can cause a non-negligible error on the simultaneous gate
operations. The error can become even more serious when the nearest-neighbor qubit-pair
approaches the frequency collision region. For example, focusing on the
qubit-pair $(Q_{0},\,Q_{1})$ in the chain system, the detuning between
transition $|0\rangle\leftrightarrow|1\rangle$ of $Q_{0}$ and
transition $|1\rangle\leftrightarrow|2\rangle$ of $Q_{1}$ (i.e., type-3 frequency
collision) takes the smallest value ($100\,\rm MHz$) among all nearby qubits, as marked (in pink shadow) in
Fig.~\ref{fig5}(b). The closer to the frequency collision region, the more serious the
cross-driving error, which can explain that simultaneous $X$ gate on qubit-pair $(Q_{0},\,Q_{1})$
has the worst gate performance in the four-qubit chain system. Similar results can also
be obtained for the qubit-pair $(Q_{0},\,Q_{3})$ in the square system.

In addition, note that as discussed
in Sec.~\ref{SecIIB1}, cross-driving induced single-qubit rotation error has been taken
into account already for the characterization of the isolated gate performance, and can lead
to leakage error since we always assume all the non-participating spectator qubit in their ground
states. However, during simultaneous single-qubit gate operations, this rotation
error occurs within the two-qubit computational subspace, thus here it does not contribute to
the leakage error but the control error for the composite gate operation. The combination
of decreased leakage error and increased control error can explain that for the
next-nearest-neighbor qubit-pair, whose detuning approaches the Type-1 frequency collision (as
discussed in Sec.~\ref{SecIIB1}, here the cross-driving effect favors single-qubit rotation error),
the ideal composite gate appears to have a little bit larger gate error than that of the simultaneous
gate operation, as shown in Fig.~\ref{fig6}(a) (we have checked this independency, and find that without
consideration of leakage error, this feature indeed disappears).

\begin{figure}[tbp]
\begin{center}
\includegraphics[keepaspectratio=true,width=\columnwidth]{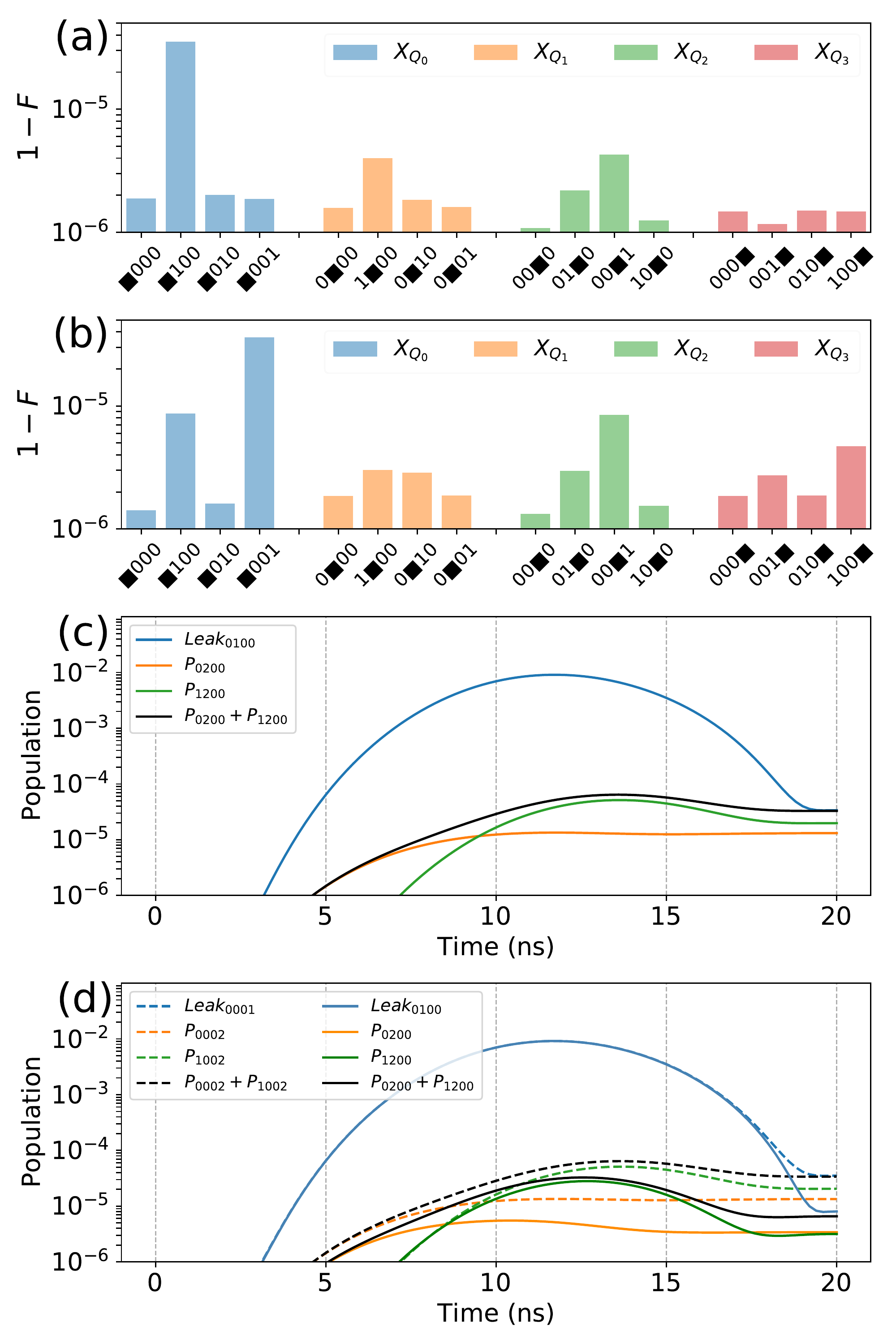}
\end{center}
\caption{$X$ gate error with the other three spectator qubits in four
different state-configurations, e.g., marked by
$\{\blacksquare 000,\,\blacksquare 100,\,\blacksquare 010,\,\blacksquare 001\}$ for $Q_{0}$. (a)
for the four-qubit chain system shown in Fig.~\ref{fig5}(b). (b) for the four-qubit square system
shown in Fig.~\ref{fig5}(c). (c) and (d) State population as a function of time
during the isolated single-qubit operations in the
four-qubit chain system and square system. Solid lines (dashed lines) denote the
result for the four-qubit system initialized in $|0100\rangle$ ($|0001\rangle$). The leakages are
defined as $Leak_{0100}=1-(P_{0100}+P_{1100})$, $Leak_{0001}=1-(P_{0001}+P_{1001})$,
where $P_{ijkl}$ represents the population in the four-qubit state $|ijkl\rangle$.}
\label{fig7}
\end{figure}

To give a more concrete illustration of the cross-driving impact on the present
four-qubit system, Figures ~\ref{fig7}(a) and ~\ref{fig7}(b) show the
X gate error with the other three spectator qubits in four different
state-configurations for the four-qubit chain and square system, respectively.
Generally, we find that compared with the isolated gate operation, gate error
is increased when a state flip of one nearby qubit occurs. Moreover, qubits
in the lower-frequency band show a more prominent decrease in gate performance,
such as $Q_{0}$ in the two four-qubit systems. This is to be expected, as detuning
between the transition $|0\rangle\leftrightarrow|1\rangle$ of the qubit in the
lower-frequency band and the transition $|1\rangle\leftrightarrow|2\rangle$
of one in the higher-frequency band, typically, $100\,\rm MHz$, is smaller than
that in the reversed case, typically, $500\,\rm MHz$. Thus, the cross-driving
effect from the low-frequency qubit to the nearby high-frequency qubit
is more serious and can result in leakage error for the low-frequency qubit.
In Figs.~\ref{fig7}(c) and~\ref{fig7}(d), we also show the time
evolution of state population during $X_{Q_{0}}$ with one of $Q_{0}$'s nearby qubit in
its excited state. For the four-qubit system initialized in $|0100\rangle$, we find
that during $X$ gate on $Q_{0}$, the qubit $Q_{1}$ state can be
excited from $|1\rangle$ to $|2\rangle$ (see Figs.~\ref{fig7}(c)
and~\ref{fig7}(d), $P_{0200}$ and $P_{1200}$), accounting for the dominated
leakage error $Leak_{0100}\equiv1-(P_{0100}+P_{1100})$. Similarly, for the four-qubit
square system prepared in $|0001\rangle$, the leakage error
$Leak_{0001}\equiv1-(P_{0001}+P_{1001})$ is mainly resulted from the population
leakage of $Q_{3}$ (see Fig.~\ref{fig7}(d), dashed line). Overall, the above analysis
shows that the dressed induced cross-driving can explain the performance
degradation of simultaneous gate operations applied to qubit-pair ($Q_{0},\,Q_{1}$)
in the four-qubit chain system and ($Q_{0},\,Q_{3}$) in the four-qubit square system.

Before leaving this subsection, we note that although the cross-driving effect
can cause almost an order of magnitude worse gate performance in the above
illustration, on average the added gate error is about $10^{-5}$,
which is an order of magnitude lower than the state-of-the-art gate
performance \cite{Chen2016,McKay2017,Somoroff2021}.
However, when qubit systems approach the frequency-collision region or have a larger
inter-qubit coupling at the idle point, this cross-driving effect can
become a dominant error source for simultaneous single-qubit gate operations.
This suggests that the cross-driving effect should be taken
seriously into account to design and construct a large multi-qubit lattice.

\begin{figure}[tbp]
\begin{center}
\includegraphics[keepaspectratio=true,width=\columnwidth]{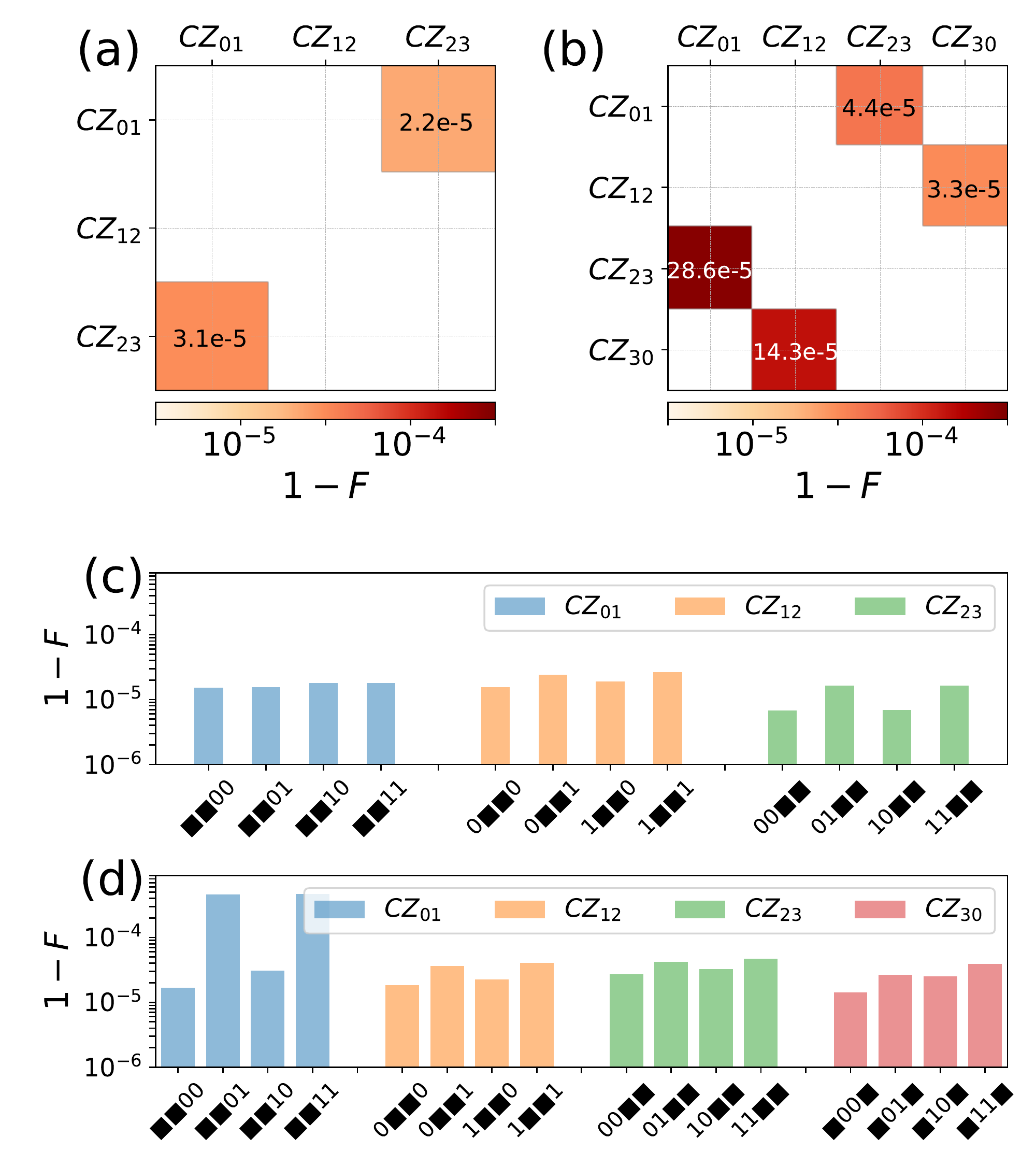}
\end{center}
\caption{Error Matrix of simultaneous CZ gates in (a) the four-qubit chain system
and (b) the four-qubit square system. (c) and (d) CZ gate error with the other two
nonparticipating qubits in four different state-configurations (i.e., $00,\,01,\,10,\,11$)
for the four-qubit chain system and the four-qubit square system.}
\label{fig8}
\end{figure}

\begin{figure*}[tbp]
\begin{center}
\includegraphics[width=15cm,height=10cm]{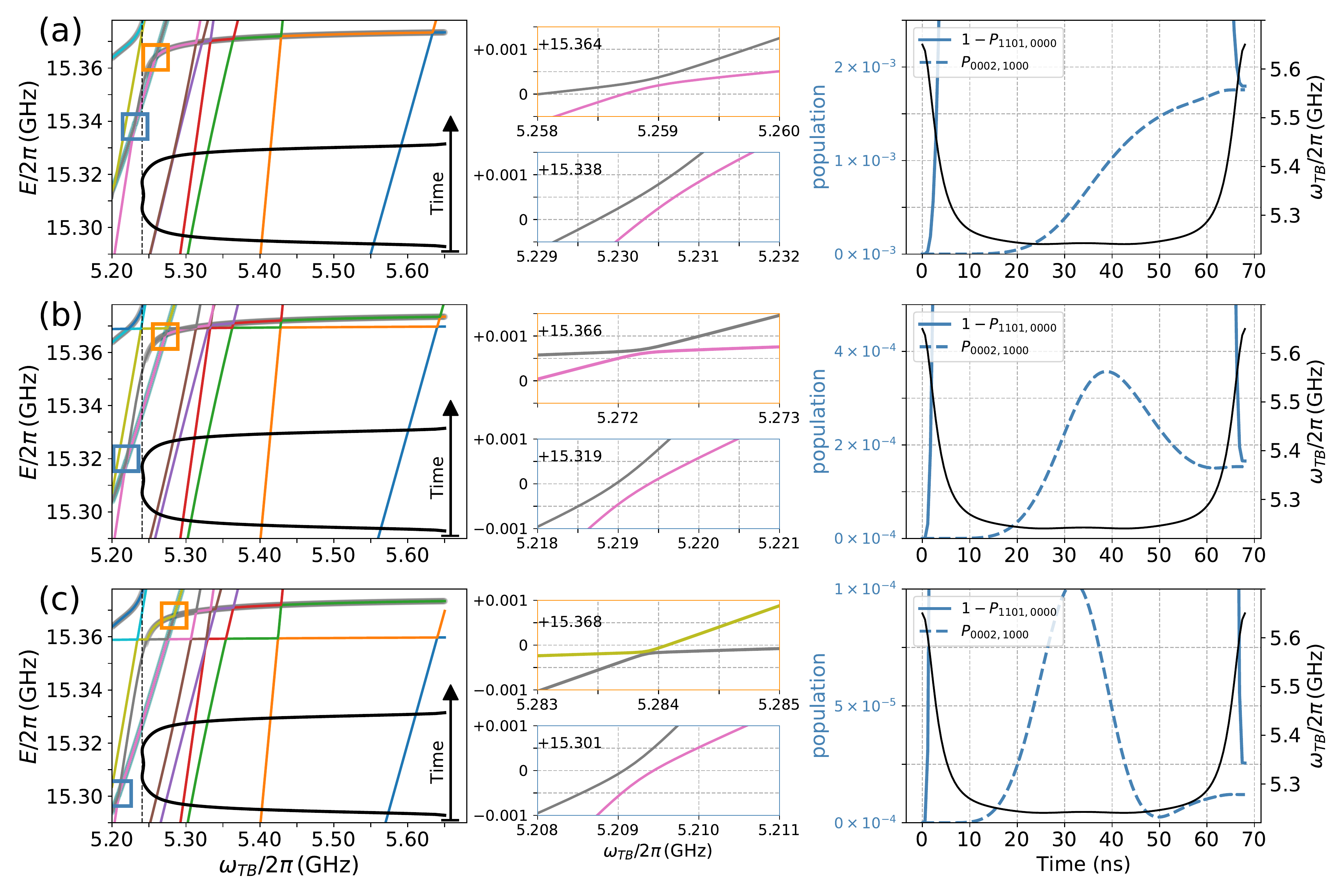}
\end{center}
\caption{ (left panel) Energy-level spectrum of the four-qubit square system shown in Fig.~\ref{fig5}.
The dashed black line indicates the working point, and the orange (steel-blue) square marks the
avoided crossing resulting from interaction $|1101,0000\rangle\leftrightarrow|0002,1000\rangle$ and $|0001,2000\rangle\leftrightarrow|0002,1000\rangle$, respectively, which are highlighted in the middle panel.
(right panel) Time evolution of state population during the implementation of $CZ_{01}$
with system prepared in the state $|1101,0000\rangle$. $P_{1101,0000}$ and $P_{0002,1000}$ denote the population on
state $|1101,0000\rangle$ and state $|0002,1000\rangle$. The black curves in (left panel)
and (right panel) represent the 68-ns shaped pulse used to realize $CZ_{01}$. (a) Qubit anharmonicities are
$\eta_{Q}/2\pi=-300\,\rm MHz$. (b) Qubit anharmonicities are $\eta_{Q}/2\pi=-310\,\rm MHz$. (c) Qubit
anharmonicities are $\eta_{Q}/2\pi=-320\,\rm MHz$. Other parameters are the same as in Fig.~\ref{fig5}(c).}
\label{fig9}
\end{figure*}

\subsection{Simultaneous two-qubit CZ gate operations}\label{SecIIIB}

In this subsection, we turn to examine the performance of simultaneous CZ gates in
the two four-qubit systems shown in Figs.~\ref{fig5}(b) and~\ref{fig5}(c). Similar to
the analysis of the single-qubit case, we consider that individual CZ gate is tuned up
and characterized with the other two nonparticipating qubits in their ground states. During
gate operations, except for the bus to which a baseband flux pulse is applied for
realizing a CZ gate, the other buses are all biased to their idle points. The CZ gate is
implemented by using a fast-adiabatic pulse given in Eq.~(\ref{eq8}) with
three Fourier coefficients $\{\lambda_{1}, \lambda_{2}, \lambda_{3}\}$, giving
rise to totally three free parameters $\{\lambda_{1}, \lambda_{2},\theta_{f}\}$.
The other parameters are: gate time $T=68\,\rm ns$, $J_{101}/2\pi=12\,\rm MHz$.
For reaching a given target error at $10^{-5}$, the three free parameters
are fixed using numerical optimization (see Appendix~\ref{A2} for details).

\subsubsection{gate performance}\label{SecIIIb1}

In Figs.~\ref{fig8}(a) and~\ref{fig8}(b), we show simultaneous CZ gate
performance in the four-qubit chain system and four-qubit square system. In the
chain system, the added error of simultaneous CZ gates is about $10^{-5}$,
giving an average added error $0.5\times10^{-5}$ for the individual CZ gate, and
in the four-qubit square system, the added error is about $2.4\times10^{-4}$
and $1.1\times10^{-4}$ for gate pairs $\{CZ_{01},\,CZ_{23}\}$ and $\{CZ_{12},\,CZ_{30}\}$,
respectively, which are about an order of magnitude larger than that of
the isolated case. Given the state-of-the-art
two-qubit gate performance, although it appears likely that the quantum crosstalk
between the gate pair $\{CZ_{01},\,CZ_{23}\}$ in the chain system
is less important, the quantum crosstalk effect on the square system can
become a near-term performance limiting factor. Thus, in the following
discussion, we can restrict our attention to the square system in which
the quantum crosstalk effect is more prominent.

To explore the exact nature of the performance
degradation, Figures ~\ref{fig8}(c) and ~\ref{fig8}(d) show the CZ gate error
with the other two nonparticipating qubits in four different
state-configurations for the chain system and square system, respectively.
One can find that except for $CZ_{01}$ in the square system, gate errors of all the CZ gates
show only a weak size dependence on the state of nonparticipating qubits. Typically,
here the added error is about $10^{-5}$. However, in the square system,
when $Q_{3}$ is in the excited state, the gate error of $CZ_{01}$ is increased by
about a factor of 28, giving rise to an added error of $4.5\times10^{-4}$.

To explain the above striking increase in the gate error, we
have examined the time evolution of the square system during
$CZ_{01}$ with $Q_{3}$ prepared in its excited state. We find that the leakage error, which is caused
by the high-order parasitic interaction among $|1101,0000\rangle$, $|0002,1000\rangle$
and $|0001,2000\rangle$, can account for the added gate error, as illustrated in Fig.~\ref{fig9}(a).
From the energy spectrum of the four-qubit square system in the left panel
of Fig.~\ref{fig9}(a), there exists two avoided crossings resulting from
interaction $|1101,0000\rangle\leftrightarrow|0002,1000\rangle$ and
$|0001,2000\rangle\leftrightarrow|0002,1000\rangle$, which are on either
side of the working point at $\omega_{\rm on}\simeq5.240\,\rm GHz$. Thus, when a
flux pulse tunes the bus to the working point and then comes back according to
the fast-adiabatic pulse given in Eq.~(\ref{eq8}), the system will first
sweep through the avoided crossing ($|1101,0000\rangle\leftrightarrow|0002,1000\rangle$)
at $5.2590\,\rm GHz$, and then approaches the avoided
crossing ($|0001,2000\rangle\leftrightarrow|0002,1000\rangle$) at $5.2305\,\rm GHz$, as
shown in the middle panel of Fig.~\ref{fig9}(a). In the right panel of Fig.~\ref{fig9}(a), where the time
evolution of state population is presented, we find that leakage to
$|0002,1000\rangle$ mainly occurs when the system approaches the avoided
crossing at $5.2305\,\rm GHz$.

Before giving a more detailed analysis of the leakage error, we note
that during an ideal fast adiabatic CZ gate operation, the qubit system should always be in the
instantaneous eigenstates of the qubit system. For the system initialized
in $|1101,0000\rangle$, the system dynamics should be restricted in the subspace
spanned by $\{|1101,0000\rangle,\,|0001,2000\rangle\}$. In light of this, the above
observation can be explained by that as the flux-pulse has a
short-time ramp (right panel of Fig.~\ref{fig9}(a), black line), and the energy gap of the avoided crossing
at $5.2590\,\rm GHz$ is about $250\,\rm kHz$, the system can thus sweep through it with a
negligible leakage to $|0002,1000\rangle$.
However, during the main part of the flux-pulse that has a flat shape, the system is almost settled
at the working point, where the instantaneous eigenstates is an almost maximal superposition
state of $|1101,0000\rangle$ and $|0001,2000\rangle$. Moreover, the working point nears the second avoided crossing
with a tiny energy gap of $500\,\rm kHz$ at $5.2305\,\rm GHz$, in which the detuning to the working
point is about $10\,\rm MHz$. Thus, an off-resonance population swap
between $|0002,1000\rangle$ and $|0001,2000\rangle$ occurs, causing significant
leakage to $|0002,1000\rangle$. Note here that during the main part of the flux-pulse,
the off-resonance population swap between $|1101,0000\rangle$ and
$|0002,1000\rangle$ can also occur due to the avoided crossing
at $5.2590\,\rm GHz$. However, as its detuning to the working point is
about $20\,\rm MHz$ and its gap is $250\,\rm kHz$, the leakage
resulting from the off-resonance interaction between $|1101,0000\rangle$ to $|0002,1000\rangle$
is expected to be far less than that from the off-resonance interaction
between $|0001,2000\rangle$ and $|0002,1000\rangle$.

The above analysis suggests that although individual gate operations with
high gate fidelity can be achieved, implementing the gate operation in
parallel can show substantial performance degradation. Moreover,
in multiqubit systems, there can exist detrimental high-order parasitic
interactions that cannot be captured by only characterizing isolated gate operations
with all other nonparticipating qubits in their ground states. On the contrary, characterizing
gate operations with other nonparticipating qubits at different state configurations
may give further insight into the quantum crosstalk behind the
performance degradation of the simultaneous gate operations \cite{Krinner2020}.

\subsubsection{mitigation of parasitic interaction involving spectator qubits}\label{SecIIIB2}

\begin{figure}[tbp]
\begin{center}
\includegraphics[keepaspectratio=true,width=\columnwidth]{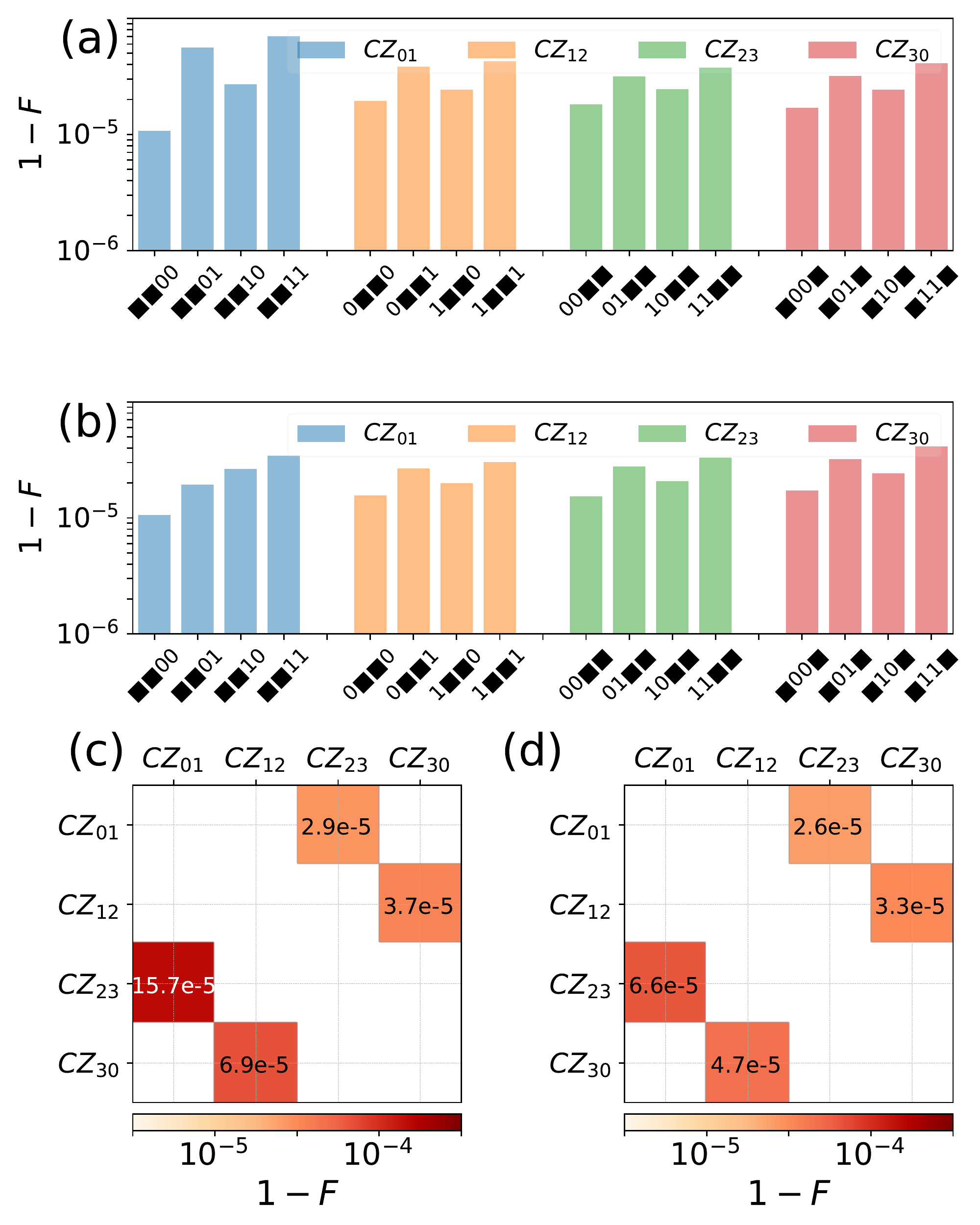}
\end{center}
\caption{(a) and (b) CZ gate error with the other two nonparticipating qubits in four different
state-configurations (i.e., $00,\,01,\,10,\,11$) for the four-qubit square system. (c)
and (d) Error matrix of the parallel implemented CZ gate operations in the four-qubit square
system. In (a) and (c), the qubit anharmonicity is $\eta_{Q}/2\pi=-310\,\rm MHz$. In (b) and (d),
the qubit anharmonicity is $\eta_{Q}/2\pi=-320\,\rm MHz$. The other
system parameters are the same as those in Fig.~\ref{fig5}(c).}
\label{fig10}
\end{figure}

From the above, the leakage to state $|0002,1000\rangle$ is the dominated error source
for $CZ_{01}$ with $Q_{3}$ in its excited state. Here, we intend to mitigate such errors
from the hardware perspective. As the leakage is mainly caused by the existence
of the parasitic interaction (i.e., $|1101,0000\rangle\leftrightarrow|0002,1000\rangle$
and $|0001,2000\rangle\leftrightarrow|0002,1000\rangle$) around the working
point, here we intend to push them away from the working point. As shown in the left
panel of Figs.~\ref{fig9}(b) and~\ref{fig9}(c), due to the increase of qubit
anharmonicities, the energy of state $|0001,2000\rangle$ is lowered, thereby, the
avoided crossings resulting from the parasitic interactions are far away from the
working point. In this way, due to the increased detuning from state $|0002,1000\rangle$,
the off-resonance population swap between state $|0001,2000\rangle$ and state $|0002,1000\rangle$
can be suppressed. Thus, as shown in the right panel of
Figs.~\ref{fig9}(b) and~\ref{fig9}(c), the leakage to $|0002,1000\rangle$ is indeed
greatly reduced, in line with expectation. Note that this is similar to the incomplete Rabi
oscillation resulting from an off-resonance driving, and the population error can be roughly
estimated by Eq.~(\ref{eq6}).

Given the suppression of the leakage error, Fig.~\ref{fig10}(a)
and Fig.~\ref{fig10}(b) show the isolated gate fidelity of $CZ_{01}$
with qubit anharmonicity $\eta_{Q}/2\pi=-310\,\rm MHz$
and qubit anharmonicity $\eta_{Q}/2\pi=-320\,\rm MHz$, respectively. The gate errors
of $CZ_{01}$ with $Q_{3}$ in its excited state are indeed greatly
decreased in both cases. In particular, for the case of $\eta_{Q}/2\pi=-320\,\rm MHz$,
there is no fundamental difference between the gate error of the CZ gate on
different qubit-pairs, and gate errors of all pairs show a weak dependence
on the state of the nonparticipating qubits. On average, the added error for
each operation is suppressed below $2\times10^{-5}$. Figures~\ref{fig10}(c)
and~\ref{fig10}(d) also present the error matrix of simultaneous CZ gates.
We find that compared to the result shown in Fig.~\ref{fig8}(b), where
$\eta_{Q}/2\pi=-300\,\rm MHz$, on average the added error of the parallel
implemented CZ gates is suppressed twofold for the square system
with $\eta_{Q}/2\pi=-310\,\rm MHz$, and sixfold for $\eta_{Q}/2\pi=-320\,\rm MHz$.

\begin{figure}[tbp]
\begin{center}
\includegraphics[keepaspectratio=true,width=\columnwidth]{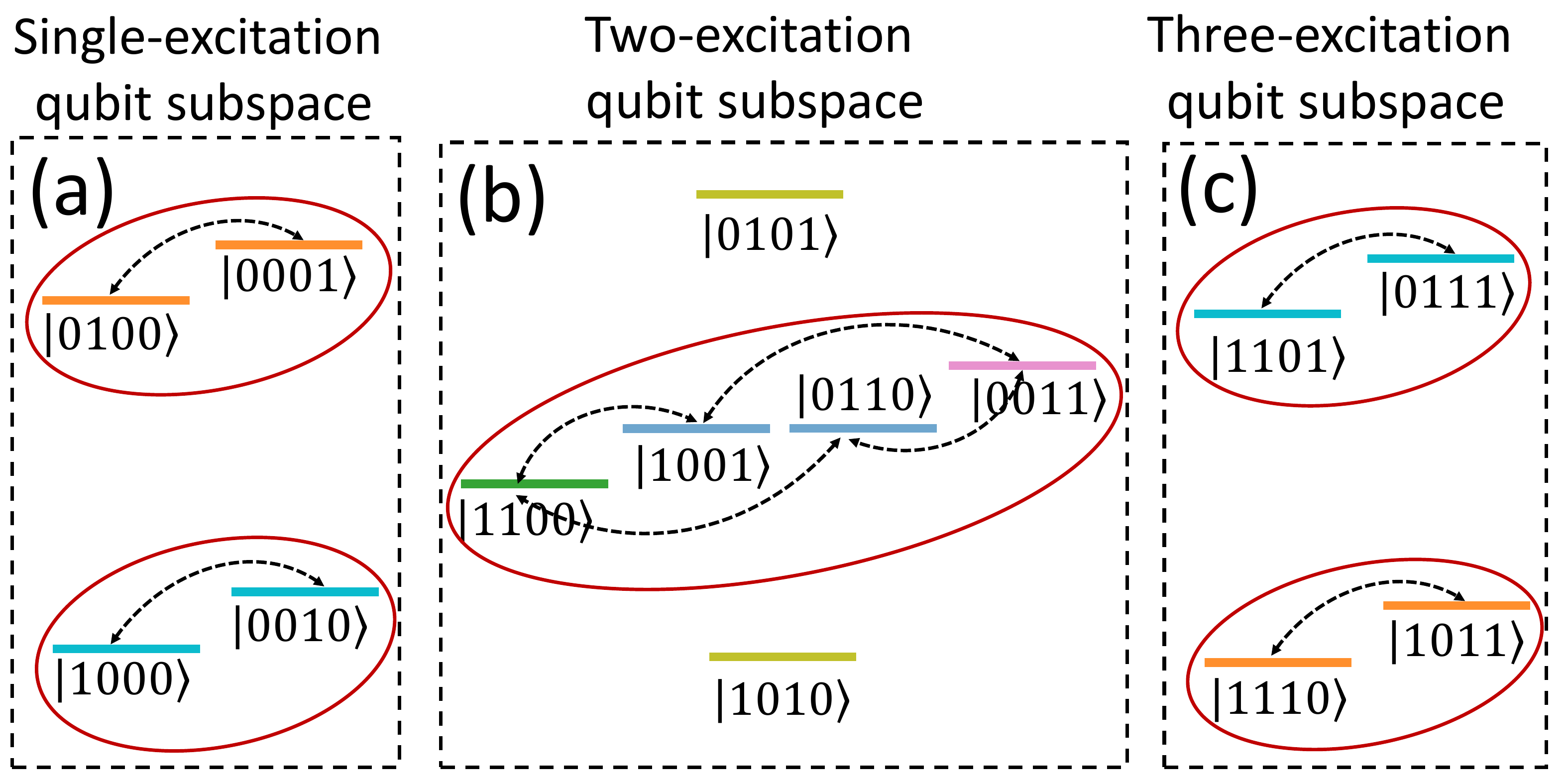}
\end{center}
\caption{Next-nearest-neighbor interaction in (a) single-excitation subspace
(b) two-excitation subspace (c) three-excitation subspace. In the
four-qubit system shown in Fig.~\ref{fig5}, the typical detuning
between next-nearest-neighbor qubits is $20\,\rm MHz$. During the CZ
gate operation, a flux pulse tunes the tunable bus from the idle point to
the working point, the bus energy levels will push down the qubit energy
levels, thus potentially diminishing the detuning between next-nearest-neighbor
qubits and leading to population swap (double-headed arrow) between
next-nearest-neighbor qubits.}
\label{fig11}
\end{figure}

Besides the leakage error, the remaining error could be attributed to the
parasitic interactions between next-nearest-neighbor qubits. As shown in
Fig.~\ref{fig11}, parasitic swap interactions (double-headed arrow)
among single-excitation, two-excitation, and three-excitation subspace
can exist through high-order processes. Since the typical detuning between next-nearest-neighbor qubits
is $20\,\rm MHz$ (see Fig.~\ref{fig5}), when a flux pulse tunes the tunable bus
from the idle point to the working point, the bus
energy levels will push down the qubit energy levels, thus potentially
diminishing the detuning between next-nearest-neighbor qubits and leading to
population swap between next-nearest-neighbor qubits. Thanks to the rather
small qubit-bus coupling, the population swap can be suppressed below
$10^{-4}$ (this was checked for the four-qubit square system prepared in
$2^{4}$ different qubit-state-configurations, see Appendix~\ref{B} for details). In
principle, this can be further improved by increasing the detuning between
next-nearest-neighbor qubits. This suggests that for systems with
increased qubit-bus or qubit-coupler coupling, as the energy level shift
is also increased, thereby, to ensure high-fidelity simultaneous gates,
a large detuning between next-nearest-neighbor qubits is needed.

\subsubsection{leakage from the avoid-crossing with a tiny gap }\label{SecIIIB3}

\begin{figure}[tbp]
\begin{center}
\includegraphics[keepaspectratio=true,width=\columnwidth]{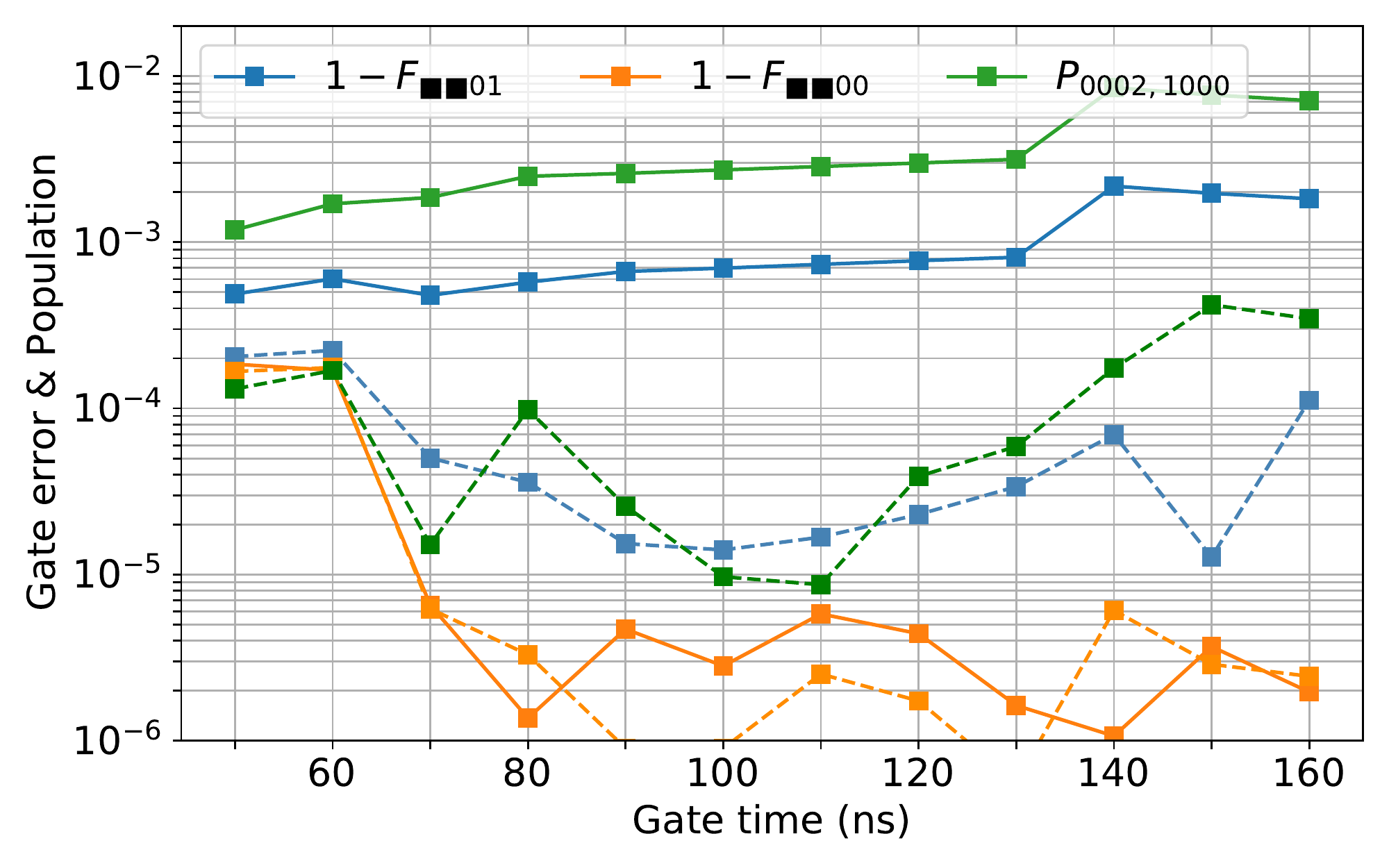}
\end{center}
\caption{Isolated $CZ_{01}$ gate error with the other two nonparticipating qubits in two different
state-configurations, i.e., $1-F_{\blacksquare\blacksquare00}$ and $1-F_{\blacksquare\blacksquare01}$,
as a function of gate time. $P_{0002,1000}$ denotes the population leakage to state $|0002,1000\rangle$
with the four-qubit square system initialized in state $|1101,0000\rangle$. Solid lines and dashed lines
denote the result with qubit anharmonicity $\eta_{Q}/2\pi=-300\,\rm MHz$ and $\eta_{Q}/2\pi=-310\,\rm MHz$,
respectively. The other system parameters used here are the same as in Fig.~\ref{fig5}(c).}
\label{fig12}
\end{figure}

We end this section by showing that without the consideration of qubit decoherence, there
could exist a performance trade-off between isolated two-qubit gate operations and simultaneous two-qubit
gate operations, i.e., in general, isolated gate operations with slower gate-speed can
suppress parasitic interactions within target qubit systems, but for simultaneous gate operations,
fast-speed gates could mitigate parasitic interactions associated with nonparticipating spectator
qubits. In Fig.~\ref{fig12}, the solid line represents the isolated $CZ_{01}$ gate error
with the other two nonparticipating qubits prepared in two different
state-configurations, i.e., $1-F_{\blacksquare\blacksquare00}$ and $1-F_{\blacksquare\blacksquare01}$,
versus gate time in the four-qubit square system shown in Fig.~\ref{fig5}(c).
In general, we find that isolated gates with slow-speed have a lower error,
but the increased gate error of $CZ_{01}$ with $Q_{3}$ in the excited state and
the leakage to state $|0002,1000\rangle$, i.e., $P_{0002,1000}$, suggest that the
gate error of simultaneous gate operations becomes serious.

\begin{figure}[tbp]
\begin{center}
\includegraphics[keepaspectratio=true,width=\columnwidth]{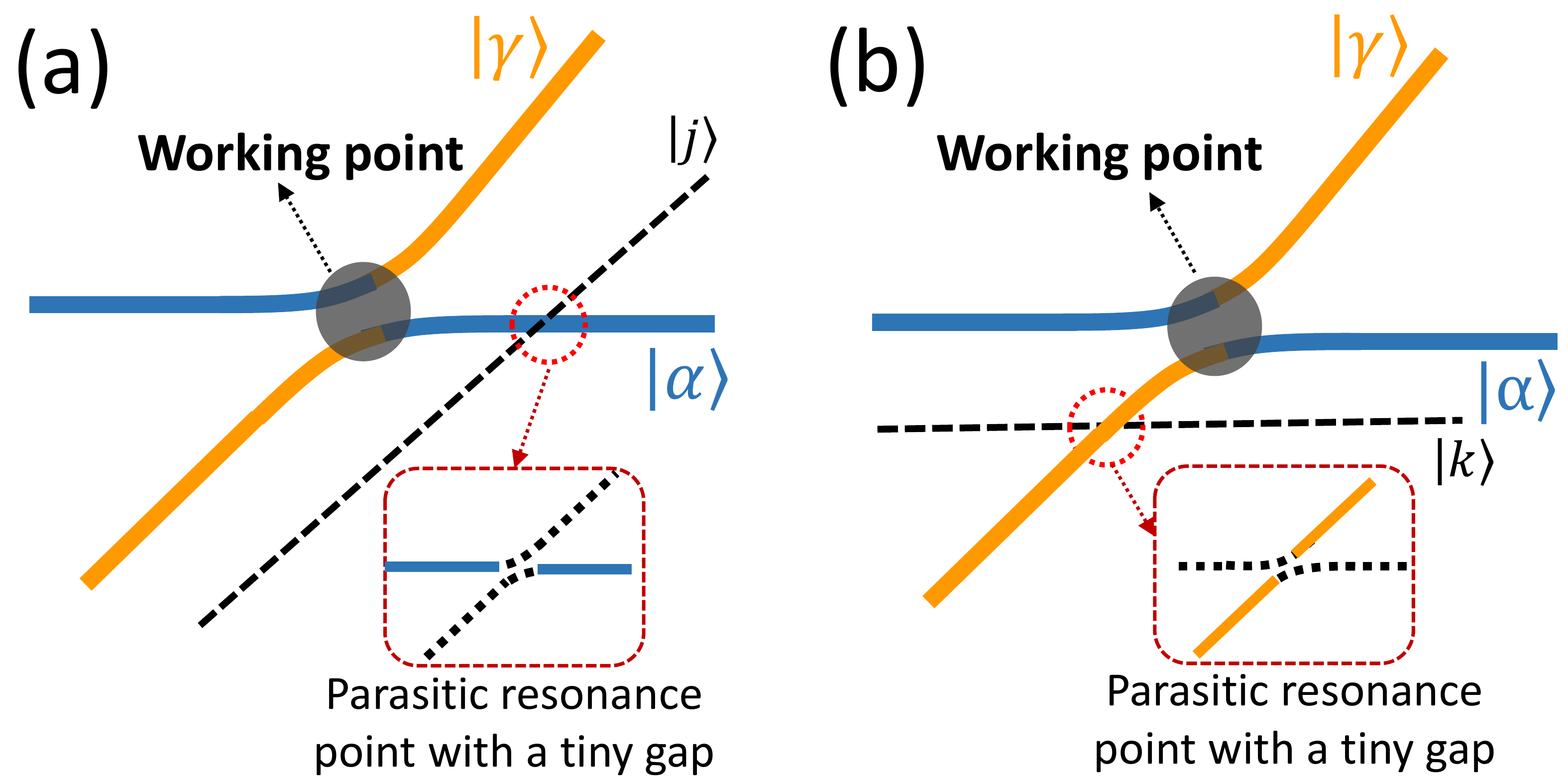}
\end{center}
\caption{Leakage channel in a multi-anticrossing system comprising a main avoided
crossing (resulting from the resonance interaction between $|\alpha\rangle$ and $|\gamma\rangle$)
with a dominated energy gap (working point)
and an avoided crossing involving (a) $|j\rangle$ or (b) $|k\rangle$ with a tiny energy gap. In (a)
during gate operations, the system will sweep through the parasitic resonance point, potentially
causing leakage to state $|j\rangle$ and in (b) the
system will approach the parasitic resonance point, resulting in a population swap between $|k\rangle$
and $|\gamma\rangle$.}
\label{fig13}
\end{figure}

To further elucidate the general physics behind the trade-off, we consider that
the present four-qubit system can be simplified to a multi-anticrossing system that
comprises a main avoided crossing with a dominated
energy gap (at the working point) and several avoided crossings with tiny energy gaps (this
is in general different from the parasitic interaction within target qubit systems, in which
the strength of the parasitic interactions is comparable to or even larger than that of
the desired interaction, thereby, the working point is chosen so that it is far detuned from
the parasitic resonance points). The
main avoided crossing results from the desired interaction used for realizing
gate operations, e.g., the interaction between $|1101,0000\rangle$ and $|0001,2000\rangle$,
and the tiny avoided crossing comes from high-order parasitic interaction, e.g., interaction
between $|1101,0000\rangle$ and $|0002,1000\rangle$, and interactions between
next-nearest-neighbor qubits, as shown in Fig.~\ref{fig11}.
In Fig.~\ref{fig13}, we show two different situations that correspond to the
two error sources discussed in Sec.~\ref{SecIIIB2}:

(i) as shown in Fig.~\ref{fig13}(a), there is a tiny avoided crossing resulting from the
interaction between $|j\rangle$ and $|\alpha\rangle$. Generally, to implement a gate
operation, one needs to bias the system to the working point. Therefore, during the
operation, the system will sweep through this tiny avoided crossing and then come back.
Thus, sweeping through this avoided crossing with fast-speed could suppress leakage to
state $|j\rangle$.

(ii) as shown in Fig.~\ref{fig13}(b), around the working point, a tiny avoided
crossing, i.e., labeled as $k$, exists, corresponding to the parasitic
interaction point (resulting from the coupling between $|\gamma\rangle$ and $|k\rangle$).
During the gate operation, the system approaches the parasitic resonance point,
thus the detuning between $|\gamma\rangle$ and $|k\rangle$ is reduced. In addition, the
energy level $|\gamma\rangle$ is pushed down due to its coupling to the state $|\alpha\rangle$,
thereby, further reducing its detuning from $|k\rangle$.  Ideally, the system dynamics should be
restricted in the subspace spanned by $\{|\alpha\rangle, \,|\gamma\rangle\}$.
However, the off-resonance interaction between the $|\gamma\rangle$ and $|k\rangle$ can result
in an incomplete population swap between the two states, especially, when the detuning
between $|\gamma\rangle$ and $|k\rangle$ becomes smaller, a population swap with a larger magnitude
will present, thus potentially causing non-negligible control or leakage error (here, i.e, leakage to
the state $|k\rangle$). Note that the same as the off-resonance error in
single-qubit systems \cite{Malekakhlagh2021}, in general,
the off-resonance swap can be suppressed by increasing the detuning between the two states,
as demonstrated in Sec.~\ref{SecIIIB2}. Furthermore, the exact
off-resonance swap error should be studied by considering the shape and the gate-time of the
control pulse \cite{Sung2021,Foxen2020,Malekakhlagh2021}. In general, pulses with a long ramp
can mitigate the off-resonance swap error.
Moreover, the error itself exhibits a periodic dependence on the gate time, and the
period generally is inversely proportional to the detuning. As shown in Fig.~\ref{fig12},
the dashed line denotes the isolated $CZ_{01}$ gate error versus gate time with an
increased qubit anharmonicity $\eta_{Q}/2\pi=-310\,\rm MHz$, giving a larger
detuning between state $|0002,1000\rangle$ and state $|0001,2000\rangle$
(Fig.~\ref{fig9}(b), about $20\,\rm MHz$).
Compared to the case of $\eta_{Q}/2\pi=-300\,\rm MHz$ (see Fig.~\ref{fig13}, solid lines),
we can find that due to the increased detuning, both the magnitude of leakage to
state $|0002,1000\rangle$ and the time of the oscillation period are decreased as expected.
Thus, to mitigate the off-resonance error, particular attention should
be paid to the parasitic resonance point (where a tiny avoided crossing exists) with a
small detuning to the working point. In this case, the smaller the detuning, the longer the
period of the error oscillation, thereby, implementing a gate operation within a
short time can mitigate the off-resonance swap error (see Fig.~\ref{fig12}, solid lines).

To circumvent the detrimental effect from these tiny-gap avoided crossings, on
the whole, we argue that implementing short gates or pushing away
these avoided-crossings could greatly relieve the population leakage and swap process
involving parasitic next-nearest-neighbor interactions. However, fast-speed gate operations
are commonly realized with potentially increased gate error within target qubit
systems, such as the leakage error \cite{Foxen2020}. In the present qubit
architecture, to suppress diabatic transitions
within the target two-qubit system, the gate-length cannot be reduced
below 50 ns, limited by the small qubit-bus coupling.
This could raise a tread-off between gate error resulting from
target qubit systems themselves and error from non-participating spectator qubits,
as shown in Fig.~\ref{fig12}. With regard to engineering quantum
systems to remove harmful tiny-gap avoided crossings, although its success
has been demonstrated in the four-qubit square system (see Sec.~\ref{SecIIIB2}),
it is still an open question as to whether this will prove useful in a large
qubit lattice.

\section{conclusion and discussion}\label{SecIV}

In this work, we first propose a qubit architecture with tunable $ZZ$ coupling, where
fixed-frequency transmon qubits are coupled via a tunable bus, and sub-100-ns CZ gates
can be realized by applying a fast-adiabatic flux-pulse to the bus. Then, we thoroughly
analyze the quantum crosstalk impact on simultaneous gate operations in this qubit architecture.
In general, we find that characterization of isolated gate errors,
especially, with all non-participating spectator qubits in their ground states, cannot
capture the full physical picture of error sources behind simultaneous gate
operations. Given this insight, we examine the isolated gate performance
with non-participating spectator qubits in various state configurations.
Combing with an inspection of the system dynamics during gate operations,
we find that to ensure high simultaneous single-qubit gate fidelities, dressing-induced
cross-driving should be seriously
considered when one operates qubits near the frequency collision regions.
For simultaneous two-qubit gates, while parasitic nearest-neighbor
interactions are commonly suppressed, parasitic next-nearest-neighbor
interactions involving spectator qubit can still exist, causing considerable
leakage or control error when qubit systems sweep through or approach these parasitic resonance
points slowly. As gate operations with slower speed can, in general, suppress unwanted interactions within
target qubit systems, but for high-order parasitic interactions involving spectator qubits,
one tends to favor short gates. This could
give rise to a trade-off between the error resulting from target qubit systems themselves and the
error from non-participating spectator qubits. Thus, our analysis suggests that in pursuit of
a functional quantum processor, the qubit architecture should be examined
carefully in the context of high-fidelity simultaneous gate operations.

To mitigate the dominated high-order parasitic interaction
in the proposed qubit architecture, we also consider
a hardware approach to mitigating their detrimental effect. We demonstrate
that by engineering qubit parameters, we can push these parasitic resonance
points far away from the working point of desired interaction. In this way,
on average, there is almost an order of magnitude improvement in gate
performance, giving the added error for simultaneous CZ gates
below $1.5\times10^{-5}$. The remaining error could be further suppressed
by increasing the detuning between next-nearest-neighbor qubits. This suggests
that the proposed qubit architecture may be a potential architecture towards
a large-scale superconducting quantum processor with low quantum crosstalk.

Although our present analysis of quantum crosstalk focuses on superconducting
qubit architecture with tunable buses, we expect that many of our conclusions may also
be applied to other qubit architecture, such as the qubit architecture with tunable
coupler \cite{Yan2018,Collodo2020,Xu2020,Foxen2020,Sete2021}, and or at the very
least, may give some physical insight into the exact nature of quantum crosstalk
in these qubit architectures. Meanwhile, the analysis procedure may
also provide a preliminary guideline for analyzing quantum crosstalk in other
qubit architecture, and may help motivate future work on bridging the gap
between high-level, hardware-agnostic crosstalk characterization \cite{Sarovar2020}
and the exact nature of crosstalk.

\begin{acknowledgments}
We would like to thank Yanwu Gu, Mengjun Hu, Yingshan Zhang, Jingning Zhang, and
Teng Ma for many helpful discussions. We also thank Yu Song and Shuang Yang for
helpful suggestions on the manuscript. This work was supported by the Beijing Natural
Science Foundation (Grant No.Z190012), the National Natural Science Foundation of
China (Grants No.11890704, No.12004042, No.11905100), the National Key Research and Development
Program of China (Grant No.2016YFA0301800), and the Key-Area Research and Development
Program of Guang Dong Province (Grant No. 2018B030326001). P.X. was supported by the Young
Fund of Jiangsu Natural Science Foundation of China (Grant No.BK20180750) and the National
Natural Science Foundation of China (Grant No.12105146).

\end{acknowledgments}

\appendix
\section{calibration procedure of gate operations}\label{A}

In this Appendix, we gave further detailed descriptions on the calibration procedure
of gate operations (i.e., single-qubit X gate and two-qubit CZ gate) discussed in
the main text. Here, for illustration purposes only, we consider the two-qubit system
shown in Fig.~\ref{fig1}. In the following discussion, we show how the single-qubit X gate
and the two-qubit CZ gate are calibrated in the two-qubit system. As in Fig.~\ref{fig1}, the
system parameters are: qubit frequency $\omega_{1(2)}/2\pi=5.000(5.200)\,\rm GHz$, anharmonicity $\eta_{1}=\eta_{2}=\eta_{\rm{TB}}=\eta\,(\eta/2\pi=-300\,\rm MHz)$, and qubit-bus
coupling $g_{1(2)}=g\,(g/2\pi=25\,\rm MHz)$ (at $\omega_{1(2)}=\omega_{\rm{TB}}=5.500\,\rm {GHz})$. In
addition, the idle point of the tunable bus is at $\omega_{TB}/2\pi=5.65\,\rm GHz$, as marked by the black
arrow in Fig.~\ref{fig1}(c). In the following discussion, notations $|Q_{1}\,Q_{2}\,TB\rangle$ represent
the full system states, and when confined to the qubit subspace, notation $|Q_{1}\,Q_{2}\rangle$ is used.

In addition, we note that the following introduced calibration procedures
are also applied to the calibration of single-qubit gates and two-qubit gates
in the four-qubit system studied in Sec.~\ref{SecIII}.

\subsection{Calibration of single-qubit X gates}\label{A1}

\begin{figure}[tbp]
\begin{center}
\includegraphics[keepaspectratio=true,width=\columnwidth]{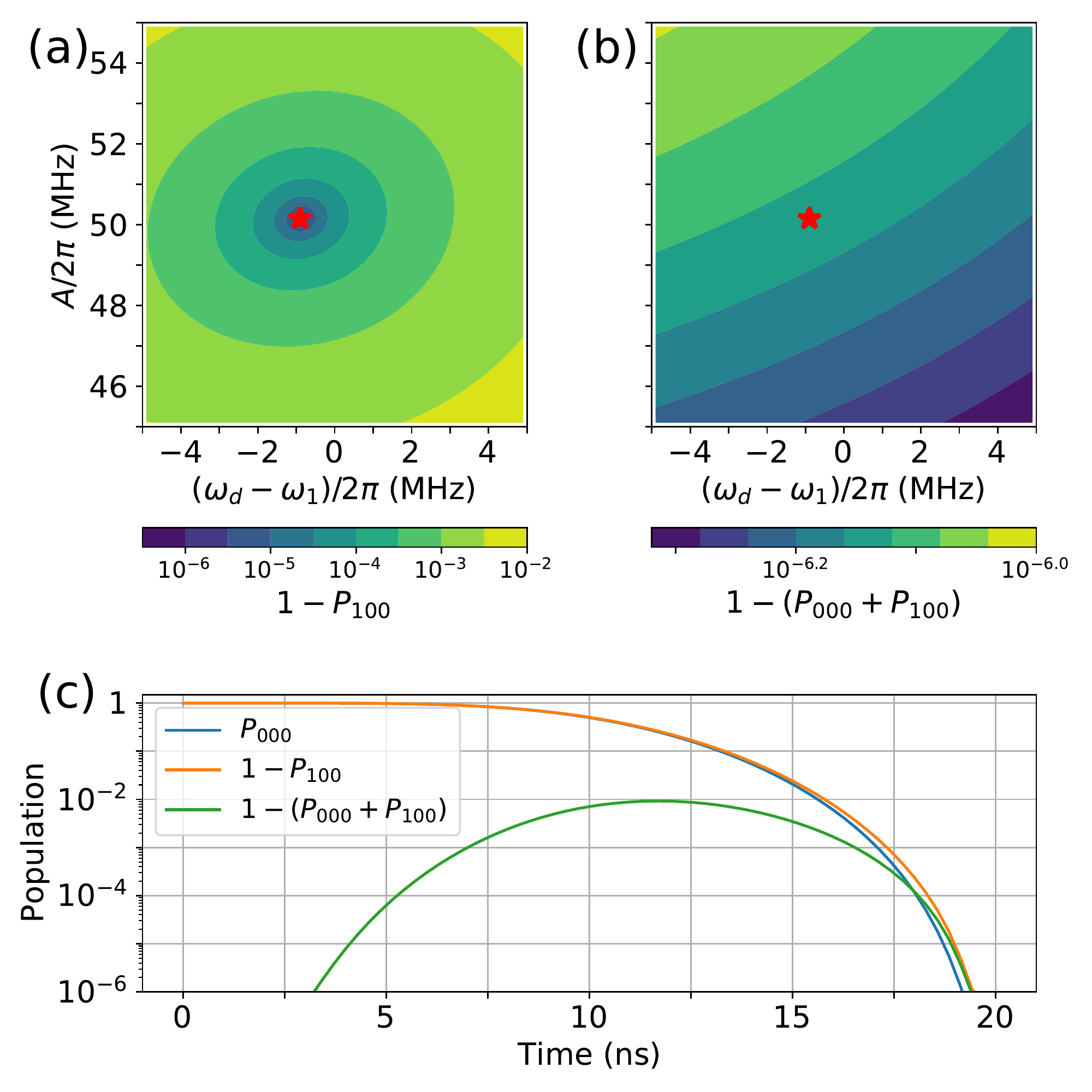}
\end{center}
\caption{Tune-up procedure for the single-qubit X gate. (a) Population
error $1-P_{100}$ as a function of the driving amplitude $A$ and the
driving detuning $(\omega_{d}-\omega_{1})$ for the two-qubit system initialized in
state $|000\rangle$, and the fixed gate length $t_{g}=20\,\rm ns$. (b) Same as in (a), instead of showing the leakage
$1-(P_{000}+P_{100})$ versus the driving amplitude $A$ and the
driving detuning $(\omega_{d}-\omega_{1})$ for the single-qubit gate operations. (c) System dynamics during the gate operation with
the optimal parameter set indicated by the red stars in (a-b) where both the population
error and the leakage are suppressed below $10^{-6}$.}
\label{fig14}
\end{figure}

As we mentioned in Sec.~\ref{SecIIB1}, in the present work, the single-qubit X gate is implemented
by using the DRAG scheme \cite{Motzoi2009,Chen2016,McKay2017}. For illustration purposes, here,
we consider the implementation of a single-qubit X gate applied to the qubit $Q_{1}$. As shown
in Eq.~\ref{eq4}, the driven Hamiltonian for implementing $X_{Q_{1}}$ is
\begin{eqnarray}
\begin{aligned}\label{eq9}
&H_{d}(t)=[\Omega_{x}(t)\cos(\omega_{d}t)+\Omega_{y}(t)\sin(\omega_{d}t)](a_{1}^{\dagger}+a_{1}),
\\&\Omega_{x}(t)=\frac{A}{2}[1-\cos(2\pi t/t_{g})],
\Omega_{y}(t)=-\frac{\alpha}{\eta}\dot{\Omega}_{x}(t),
\end{aligned}
\end{eqnarray}

The $X_{Q_{1}}$ gate is tuned up by adjusting the driving amplitude $A$ and the driving detuning of
the pulse from the qubit bare frequency $(\omega_{d}-\omega_{1})$ for a fixed gate length ($t_{g}=20\,\rm ns$)
and the parameter $\alpha=0.5$. Figure~\ref{fig14}(a) shows the population error $1-P_{100}$ ($P_{100}$
denotes the population in state $|100\rangle$ at the end of the gate operation) as a function
of the driving amplitude $A$ and the driving detuning $(\omega_{d}-\omega_{1})$ for the two-qubit system
initialized in state $|000\rangle$. In Fig.~\ref{fig14}(b), we also show the leakage
error $1-(P_{000}+P_{100})$. One can find that at the minimal error
point, i.e., $\{A/2\pi\simeq50.15\,{\rm MHz}, (\omega_{d}-\omega_{1})/(2\pi)\simeq0.9\,\rm MHz\}$,
marked by the red stars in Figs.~\ref{fig14}(a) and (b), both the population error and leakage
is suppressed below $10^{-6}$, as shown in Fig.~\ref{fig14}(c). Note here that in the present setting,
as expected, the obtained optimal driving frequency is indeed at the dressed qubit frequency of $Q_{1}$.

After obtaining the optimal parameter set from the result shown in Fig.~\ref{fig14}, we
characterize the fidelity of $X_{Q_{1}}$ gate by using the metric given in Eq.~\ref{eq7}, i.e.,
\begin{equation}
\begin{aligned}
F(\tilde{U}_{X},X)=\frac{{\rm Tr}(\tilde{U}^{\dagger}_{X}\tilde{U}_{X})+|{\rm Tr}(X^{\dag}\tilde{U}_{X})|^{2}}{6},
\end{aligned}
\end{equation}
where $\tilde{U}_{X}$ denote the actual implemented X gate operation, and
\begin{equation}
X=\left(
\begin{array}{cc}
0 & 1 \\
1 & 0 \\
\end{array}
\right).
\end{equation}

According to the two-qubit system Hamiltonian given in Eq.~\ref{eq1} and the driven Hamiltonian
given in Eq.~\ref{eq9}, the actual evolution operator is
\begin{equation}
\begin{aligned}
U_{1}=\mathcal{\hat{T}} {\rm exp}\left(-i\int_{0}^{t_{g}}H_{X}(t)dt\right),
\end{aligned}
\end{equation}
where $H_{X}(t)=H+H_{d}(t)$, and $\mathcal{\hat{T}}$ denotes the time-ordering
operator. Thus, in the single-qubit computational subspace for $Q_{1}$, the actual
implemented X gate operation is given as
\begin{equation}
\begin{aligned}
\tilde{U}_{X}=\mathcal{P}_{1}U_{1}\mathcal{P}^{\dagger}_{1},
\end{aligned}
\end{equation}
where $\mathcal{P}_{1}$ is the projected operator defined in the
computational subspace of the full system, i.e., the subspace
spanned by $\{|000\rangle,\,|100\rangle\}$. Finally, to account for the
local single-qubit phase \cite{Chen2016, McKay2017}, the gate fidelity
is obtained as \cite{Zhao2020b, Zhao2020}
\begin{equation}
\begin{aligned}
F={\rm maximize}_{\{\phi\}}\,\,F(U_{\rm{post}1}(\phi)\tilde{U}_{X},X),
\end{aligned}
\end{equation}
with $U_{\rm{post}1}(\phi)=e^{-i\phi Z/2}$, where Z represents the single-qubit Pauli
operator $\sigma_{z}$.

\subsection{Calibration of CZ gates}\label{A2}

As mentioned in Sec.~\ref{SecIIB2} and Sec.~~\ref{SecIIIB}, the CZ gate
\begin{equation}
CZ=\left(
\begin{array}{cccc}
1 & 0 & 0& 0 \\
0 & 1 & 0 & 0 \\
0 & 0 & 1 & 0\\
0 & 0 & 0 & -1\\
\end{array}
\right)
\end{equation}
is realized by applying a fixed-length ($T=68\,\rm ns$) flux pulse to the tunable
bus, tuning its frequency from its idle point ($\theta_{0}$) to the working point ($\theta_{f}$) and
then coming back. The pulse shape is given in Eq.~\ref{eq8} with three Fourier
coefficients $\{\lambda_{1},\lambda_{2},\lambda_{3}\}$, where
$\lambda_{1}+\lambda_{3}=1$. Thus, this gives rise to totally three free
parameters $\{\lambda_{1},\lambda_{2},\theta_{f}\}$. For reaching a target
error at around $10^{-5}$, the free parameter set is determined by optimizing the CZ gate
fidelity according to the metric given in Eq.~\ref{eq7}, i.e.,
\begin{equation}
\begin{aligned}
F(\tilde{U}_{cz},CZ)=\frac{{\rm Tr}(\tilde{U}^{\dagger}_{cz}\tilde{U}_{cz})+|{\rm Tr}(CZ^{\dag}\tilde{U}_{cz})|^{2}}{20}.
\end{aligned}
\end{equation}

Similar to the case of single-qubit gates, according to the system Hamiltonian given in Eq.~\ref{eq1} and the
pulse shape given in Eq.~\ref{eq8}, the actual evolution operator for the gate operation is
\begin{equation}
\begin{aligned}
U_{2}=\mathcal{\hat{T}} {\rm exp}\left(-i\int_{0}^{T}H(t)dt\right).
\end{aligned}
\end{equation}
Thus, truanted to the two-qubit computational subspace, the actual implemented CZ gate operation
is given as
\begin{equation}
\begin{aligned}
\tilde{U}_{cz}=\mathcal{P}_{2}U_{2}\mathcal{P}^{\dagger}_{2},
\end{aligned}
\end{equation}
where $\mathcal{P}_{2}$ is the projected operator defined in the two-qubit subspace spanned by $\{|000\rangle,\,|010\rangle,\,|100\rangle,\,|110\rangle\}$. Finally, to account for the
local single-qubit phases \cite{Zhao2020b, Zhao2020}, the gate fidelity
is obtained as
\begin{equation}
\begin{aligned}
F={\rm maximize}_{\{\phi_{1},\phi_{2}\}}\,\,F(U_{\rm{post}2}(\phi_{1},\phi_{2})\tilde{U}_{cz},CZ),
\end{aligned}
\end{equation}
with $U_{\rm{post}2}(\phi_{1},\phi_{2})=e^{-i\phi_{1} ZI/2}e^{-i\phi_{2} IZ/2}$,
where Z and I represent the single-qubit Pauli operator $\sigma_{z}$ and identity
operators, and the order indexes the qubit number.

\begin{figure*}[tbp]
\begin{center}
\includegraphics[width=16cm,height=11cm]{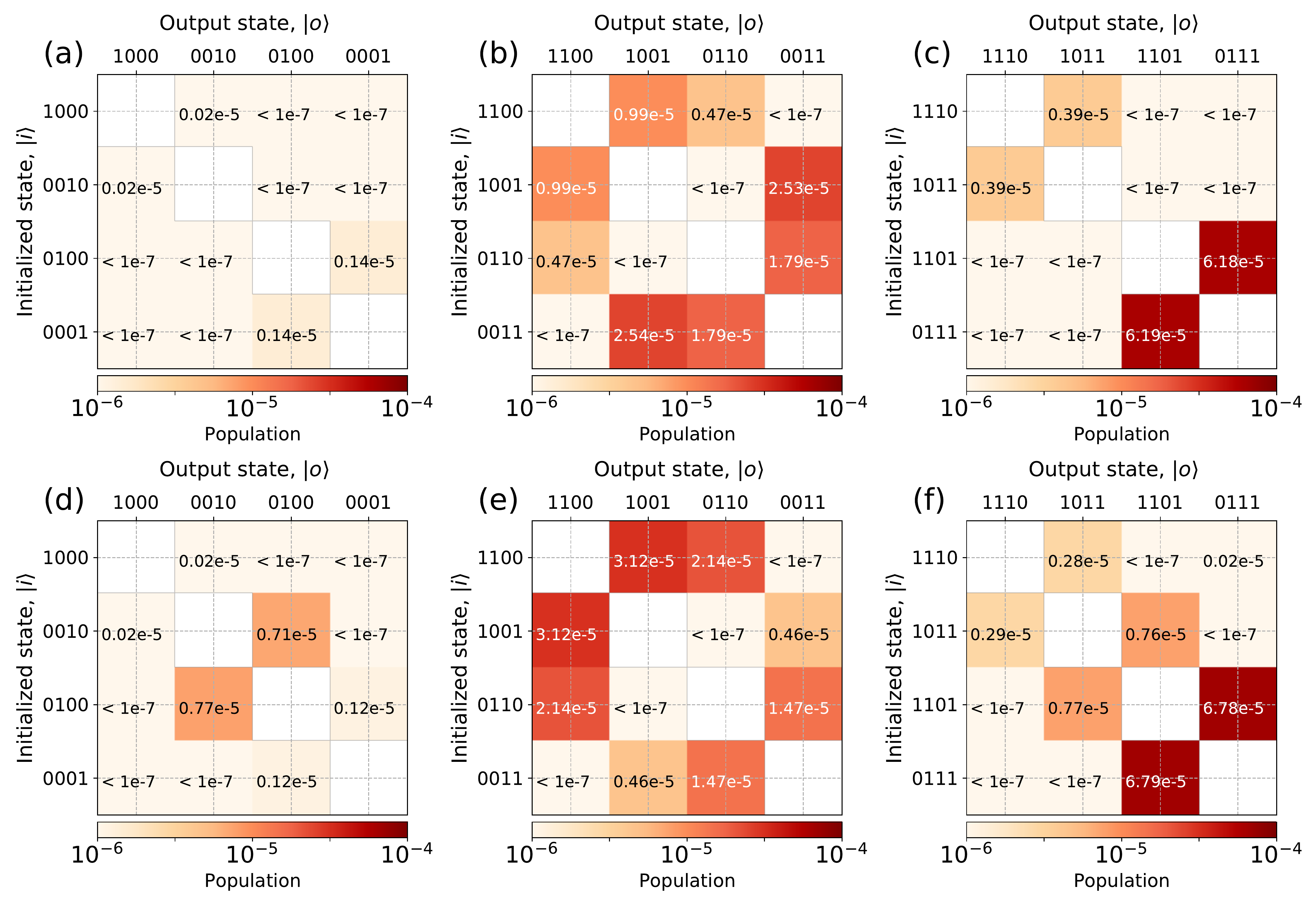}
\end{center}
\caption{Population swap matrix $P(o|i)$ for the implementation of simultaneous CZ gates in
the four-qubit square system shown in Fig.~\ref{fig5}(c). The matrix element $P(o|i)$ denotes the
population in state $|o\rangle$ for the qubit system initialized in
state $|i\rangle$. [(a),(d)] for the qubit system initialized in the single-excitation
subspace $\{|1000\rangle,\,|0010\rangle,\,|0100\rangle,\,|0001\rangle\}$, [(b),(e)] for
the two-excitation subspace $\{|1100\rangle,\,|1001\rangle,\,|0110\rangle,\,|0011\rangle\}$, and [(c),(f)] for
the three-excitation subspace $\{|1110\rangle,\,|1011\rangle,\,|1101\rangle,\,|0111\rangle\}$. (a-c) Population
swap matrix for the parallel implementation of $CZ_{01}$ and $CZ_{23}$.
(d-f) Population swap matrix for the parallel implementation of $CZ_{12}$ and $CZ_{30}$.}
\label{fig15}
\end{figure*}

\section{population swap between next-nearest-neighboring qubits}\label{B}

As shown in Fig.~\ref{fig11} and mentioned in Sec.~\ref{SecIIB2}, the population swap between
next-nearest-neighbor qubit pairs which results from off-resonance interaction among them
in the four-qubit system, can contribute to the gate error of simultaneous CZ gates.
Figure~\ref{fig15} shows the population swap matrix $P(o|i)$ in the single-excitation
subspace $\{|1000\rangle,\,|0010\rangle,\,|0100\rangle,\,|0001\rangle\}$, the two-excitation subspace $\{|1100\rangle,\,|1001\rangle,\,|0110\rangle,\,|0011\rangle\}$, and the three-excitation
subspace $\{|1110\rangle,\,|1011\rangle,\,|1101\rangle,\,|0111\rangle\}$ of the four-qubit square
system shown in Fig.~\ref{fig5}(c). Each matrix element $P(o|i)$ denotes the population in
state $|o\rangle$ after the gate operation with the qubit system initialized in
state $|i\rangle$.

From Fig.~\ref{fig15}, one can find that a significant population swap only exists for
the state transition process involving next-nearest-neighboring qubits. These transitions are
indicated by double-headed arrows connecting the states of next-nearest-neighboring qubits, as
shown in Fig.~\ref{fig11}. Moreover, as these population swaps are enabled by high-order
processes, the population swap errors are suppressed below $10^{-4}$.

\end{document}